\documentclass[10pt,twocolumn,aps,preprintnumbers,amsmath,amssymb,footinbib,superscriptaddress,longbibliography,floatfix,pre]{revtex4-2}

\usepackage{graphicx}% Include figure files
\usepackage{bm}% bold math
\usepackage{dsfont}
\usepackage[dvipsnames]{xcolor}
\usepackage{tikz}
\usepackage{verbatim}
\usepackage{units}
\usepackage{multirow}
\usepackage{mathrsfs}
\usepackage{leftidx}
\usepackage{xspace}
\usepackage{diffcoeff}  % \dl x -> dx, \diff[n]{y}{x}[x_0] -> \frac{d^n y}{dx^n}|_{x_0}, \diff*{y}{x} -> \frac{d^n}{dx^n}y, \diffp for partial derivatives
\usepackage{bbm}
\usepackage{mathtools}

\usepackage{array}
\usepackage{tabularx}
\usepackage[normalem]{ulem}
\usepackage{bbold}
\usepackage{pifont}
\usepackage[utf8]{inputenc}
\usepackage[english]{babel}

\usepackage{hyperref}
\hypersetup{
    colorlinks=true,
    linktoc=all,
    linkcolor=blue,
    breaklinks=true,
    citecolor=blue,
    urlcolor=blue
}

\definecolor{darkgreen}{RGB}{6, 153, 38}

\addto\captionsenglish{}

\newcommand{\beq}{\begin{equation}}
\newcommand{\eeq}{\end{equation}}

\DeclarePairedDelimiter{\bra}{\langle}{\rvert}
\DeclarePairedDelimiter{\ket}{\lvert}{\rangle}
\DeclarePairedDelimiterX\braket[2]{\langle}{\rangle}{#1\,\delimsize\vert\,\mathopen{}#2}
\DeclarePairedDelimiterX\ketbra[2]{\lvert}{\rvert}{#1\,\delimsize\rangle\mathopen{}\delimsize\langle\,\mathopen{}#2}

\DeclarePairedDelimiterX\Braket[2]{(}{)}{#1\,\delimsize\vert\,\mathopen{}#2}
\DeclarePairedDelimiterX\Ketbra[2]{\lvert}{\rvert}{#1\,\delimsize)\mathopen{}\delimsize(\,\mathopen{}#2}

\begin{document}

\title{Stochastic and quantum autonomous information machine:\texorpdfstring{\\}{ }
efficiency fluctuations, precision and optimization}
\author{Lucas Trigal}
\affiliation{Instituto de F\'isica Interdisciplinar y Sistemas Complejos IFISC (CSIC-UIB), E-07122 Palma de Mallorca, Spain\looseness=-1}

\author{Gonzalo Manzano}
%\date{\today}
\affiliation{Instituto de F\'isica Interdisciplinar y Sistemas Complejos IFISC (CSIC-UIB), E-07122 Palma de Mallorca, Spain\looseness=-1}
\begin{abstract}
We investigate the conversion between thermal and informational resources in an autonomous machine coupled to thermal reservoirs and a tape, playing the role of an information reservoir. The machine is able to perform information erasure driven by heat, and information-powered refrigeration, for which we introduce suitable efficiency definitions. We focus on the role of fluctuations in the currents and stochastic efficiencies, as well as in operation at optimal performance. The model is also extended to allow a coherent information reservoir, which is able to modify operation and performance. All in all, our work provides insights for the design and optimization of information-to-work conversion devices at small scales.
\end{abstract}

\maketitle

\section{Introduction}
\label{sec:intro}

Maxwell's demon though experiment and Szilard's engine illustrated how information about the microscopic state of a system can be exploited to perform an otherwise impossible thermodynamic task~\cite{LeffRex}. On the other side, Landauer's principle establishes the complementary cost associated with information processing and erasure~\cite{Landauer1961,bennett1982thermodynamics}. These ideas form the basis of information thermodynamics, where energetic and informational resources are treated on the same footing within a common nonequilibrium statistical mechanics framework~\cite{Parrondo2015}. In this setting, the information that may be acquired to allow a feedback operation can be quantified, and explicitly incorporated into the second law~\cite{Sagawa2008,Horowitz_2011}. Similarly, this allows a quantification of the work required to restore a memory storing the information back to its initial state~\cite{Sagawa2009}, even when this operation is done at finite time~\cite{Zulkowski2014,Proesmans2020}. These concepts have been demonstrated experimentally in a number of setups, including colloidal particles~\cite{Toyabe2010,Berut2012,jun2014high}, nanoelectronic systems~\cite{Koski2014,koski2014b,Scandi2022}, or untracold atoms~\cite{Yam2018,aimet2025experimentally}.

A central question since the inception of Maxwell's demon is how such information-to-energy conversion processes can be implemented autonomously, without externally prescribed sequences of measurements and feedback operations~\cite{smoluchowski1927experimentell,feynman1965feynman,astumian1998fluctuation}. Autonomous information machines incorporate the information-processing mechanism directly into their time-independent dynamics~\cite{serreli2007molecular,Strasberg2013,Horowitz2013,Horowitz2014,Koski2015,Shiraishi2015,Ptaszy2018,Sanchez2019,Bhattacharyya2022}. A paradigmatic realization consists of coupling a physical system to an information reservoir~\cite{Deffner2013,Barato_2014}, typically represented by a stream of bits that sequentially interact with the system. By conditioning transitions in the system, their statistical state can act as a thermodynamic resource, even if no energy is exchanged with the device~\cite{Freitas2021}. Prominent examples of this mechanism allowing work extraction, or heat transport against a temperature gradient, have been proposed using simple stochastic models~\cite{Mandal2012,Mandal2013,Barato_2013} (sometimes refereed to as information ratchets), while analogous quantum versions have been proposed e.g. in Refs.~\cite{Deffner2013q,Chapman2015,manzano2018entropy}. These models have further inspired extensions performing more complex operations on the tape, thus leading to thermodynamic models of computing machines~\cite{Boyd2018,Loomis2020,Jurgens2020} (see also Ref.~\cite{wolpert2019stochastic}).\looseness=-1

Although the interconversion of information and energy in autonomous stochastic machines is nowadays well understood at the average level, the role of fluctuations in the performance of these devices have attracted interest only very recently~\cite{Monsel2025,Monsel2026} and remain largely unexplored despite being of fundamental importance at micro and nanoscales~\cite{Barato2015,Gingrich2016, Horowitz2020}. Indeed, fluctuations may severely affect the precision of the currents and hence the thermodynamic operation in small thermal machines~\cite{Campisi_2015,Pietzonka_2018,
Miller_2020,Saryal_2021}. The correlated behavior of the fluctuations in different currents may also impact the efficiency of the machine, which can also be defined at the stochastic level~\cite{Verley2014a}. In this context, a general theory of efficiency fluctuations has been developed~\cite{Verley2014b,Manikandan2019} which has been not tested before in informational devices. Moreover, the potential beneficial or detrimental role of quantum effects in the operation of informational machines is still an open question~\cite{Ptaszy2019,Bernal2026}.

In this work, we contribute to fill this gap by proposing and studying a solvable model of an autonomous informational thermal machine, where fluctuations in the currents and efficiency can be easily accessed owing to the tight coupling between all energetic and informational currents. The machine is inspired in elementary models for autonomous quantum thermal machines~\cite{Brunner2012, Manzano2019,Hammam2021}(for a review see \cite{Mitchison2019}). It interacts with two thermal reservoirs at different temperatures, and with an information reservoir, modeled as an stream of bits in a tape that exchange information, but no energy, with the machine. The machine is designed complementarily perform information erasure powered by a heat flux, or information-assisted cooling of a cold thermal reservoir, in analogy to the Maxwell's refrigerator model in Ref.~\cite{Mandal2013}. By defining suitable efficiencies at the macroscopic and stochastic level, we also evaluate and optimize the performance of the informational machine in their different regimes of operation, with special emphasis in the shape of the fluctuations and precision of the currents. Finally, we explore a quantum extension of the model where the information reservoir is allowed to bear quantum coherence, to determine its impact on the operating regimes and overall performance of the machine. 

The article is organized as follows: In Sec.\ref{sec:informational_machine}, we introduce the model of the autonomous informational machine and discuss its basic functioning. The stationary thermodynamic currents at average and fluctuating levels are characterized in Sec.\ref{NESS_thermo}, together with their entropy balance. Section \ref{sec:thermodynamic_regimes} is devoted to discuss the main operational regimes of the machine and the precision of the currents. In Sec. \ref{sec:efficiencies}, we introduce suitable informational efficiencies and investigate their stochastic fluctuations. Section \ref{sec:Optimization} addresses the optimization of the performance in the different regimes and Sec.\ref{sec:coherence} discusses the effects of coherence in a quantum extension of the machine. Finally, Sec.\ref{sec:conclusion} summarizes our main findings and discusses possible implementations.

\section{Informational thermal machine}
\label{sec:informational_machine}
\begin{figure}[t]
    \centering
    \includegraphics[width=\columnwidth]{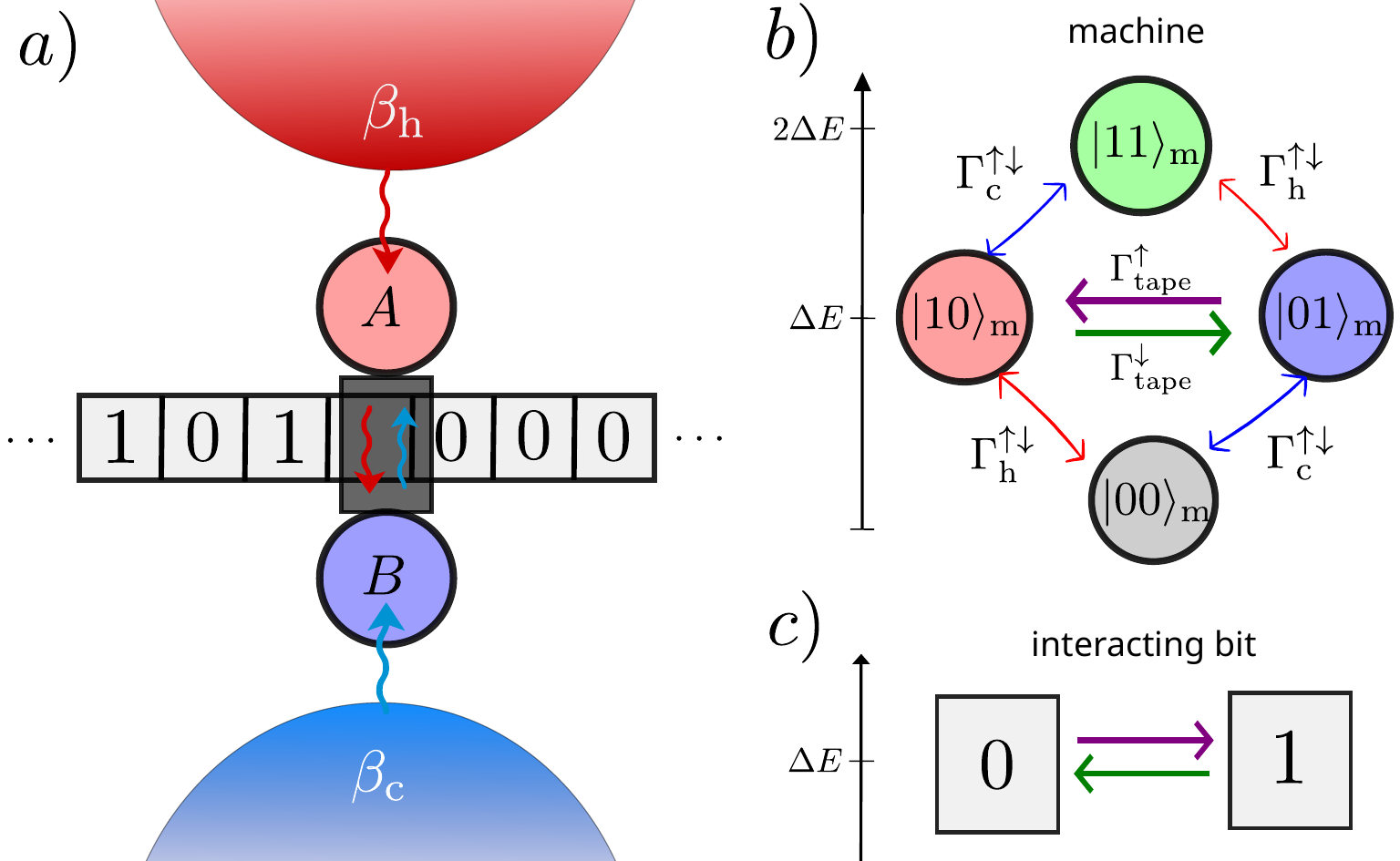}
    \caption{a) Sketch of the informational machine, composed by two-states subsystems $A$ and $B$, which exchanges heat with hot (top) and cold (bottom) thermal baths. The machine interacts with the bits of an informational reservoir (center tape) one at a time (shadow area). b) State diagram of the machine. Transitions are driven by the hot reservoir (red arrows), cold reservoir (blue arrows) and the interaction with the tape (purple and green arrows) are highlighted with their respective rates. c) Transitions between degenerate states of the machine operate bit flips on the tape bits (see corresponding colors) at same energies.}
    \label{fig:ModelD}
\end{figure}

We consider a stochastic informational machine driven by two thermal baths acting on a tape of bits (a skectch is depicted in Fig.~\ref{fig:ModelD}a). This model is inspired by autonomous quantum thermal machines~\cite{Brunner2012} and specially by the setups introduced in Refs.~\cite{Manzano2019,Hammam2021}. The machine consists of a couple of identical two-level subsystems, $A$ and $B$, each with an energy difference of $\Delta E$. The state space of the machine is therefore composed of four configurations, which we denote using a bra-ket notation~\footnote{Here the bra-ket notation is used only as a compact label for the classical states of the system, while it is upgraded to denote quantum states in Hilbert space in Sec.~\ref{sec:coherence}.} as $\ket{i \, j}_{\rm m}$ 
with $i,j = \{0,1\}$ and where the subscript m stands for ``machine". The state of the machine is fully specified by the occupation probabilities of these four states, which can be compactly included in a vector $\bm{p}_{\rm m} = (p_{00},p_{01},p_{10},p_{11})^T$, with corresponding energies given by $\bm{H}_{\rm m} = (0, \Delta E, \Delta E, 2\Delta E)^T$ (see Fig.~~\ref{fig:ModelD}b). Notice that, due to the identical energy spacing of the two subsystems, the composite states of the machine $\ket{0 1}_{\rm m}$ and $\ket{1 0}_{\rm m}$ are degenerate in energy, leading to a degenerate subspace.

Each subsystem is locally coupled to a thermal reservoir (or bath) at a different inverse temperature, $\beta_{\rm c}$ and $\beta_{\rm h}$, corresponding to cold and hot temperatures respectively, i.e. $\beta_{\rm c} \geq \beta_{\rm h}$. These thermal baths induce excitation and relaxation processes leading to jumps over the machine states with rates denoted $\Gamma_i^{\uparrow}$ and $\Gamma_i^{\downarrow}$, respectively, which satisfy the local detailed balance condition,
\begin{equation} \label{eq:rates}
{\Gamma_i^{\uparrow}}/{\Gamma_i^{\downarrow}}=e^{-\beta_i \Delta E},
\qquad i\in\{c,h\}.
\end{equation}
As a consequence, in the absence of additional couplings or interactions, each machine subsystem relaxes to a local equilibrium distribution associated with its own reservoir temperature. In that situation, the composed distribution of the machine reads $\bm{p}_{\rm m}^{\rm eq} = (p_{00}^{\rm eq}, p_{01}^{\rm eq}, p_{10}^{\rm eq}, p_{11}^{\rm eq})^T =(Z_{\rm h} Z_{\rm c})^{-1}(1, e^{-\beta_{\rm c} \Delta E}, e^{-\beta_{\rm h} \Delta E}, e^{-(\beta_{\rm h} + \beta_{\rm c})\Delta E})^T$, where we introduced the local partition functions $Z_i= 1 + e^{-\beta_i \Delta E}$ for each machine subsystem, $i={\rm h, \rm c}$.

In order to perform informational operations, the machine couples to a so-called information reservoir, namely a device which is able to exchange information with the machine, but not energy~\cite{Deffner2013,Barato_2014}. Such information reservoir may represent a memory device, and it is modeled here as a tape of bits, i.e. a sequence of two-states systems labeled as $0$'s or $1$'s. The machine acts over the tape by producing probabilistic bit flips through short interactions between the machine and the tape which occur at Poisson-distributed times (see Fig.~\ref{fig:ModelD}c). It operates over one single bit at a time and, after each interaction, the tape moves one position forward so that it can act over the next bit of the tape in the next interaction.  
We assume that the two-state systems representing the bits in the tape are also energetically degenerate~\footnote{Technically, we ascribe the same energy $\Delta E$ to both states $\ket{0}_{\rm tape}$ and $\ket{1}_{\rm tape}$ of the tape bits}, and that each interaction with the tape involves always a ``fresh" bit in the tape (that did not interacted before with the machine), prepared according to the same fixed initial distribution $\bm{p}_{\rm tape} = (p_0^{\rm tape}, p_1^{\rm tape})^T$ with $p_1^{\rm tape}=1-p_0^{\rm tape}$. This type of interaction can be viewed as a classical version of some quantum collisional models~\cite{Manzano2019,Hammam2021,Ciccarello2022}. Similar sequential interactions have been also considered in models for autonomous Maxwell demons~\cite{Mandal2012,Mandal2013} and information ratchets~\cite{Boyd2018,Loomis2020,Jurgens2020}. 

The stochastic evolution of the machine subjected to weak interaction with both thermal reservoirs and the tape is governed by a continuous-time master equation $\dot{\mathbf{p}}_{\rm m}=\mathbb{W} \, \mathbf{p}_{\rm m}$, where $\mathbb{W}$ is the rate matrix containing contributions from the hot and cold baths, and the informational tape (see App.~\ref{app:classical_generator} for details). 
While the thermal reservoirs induce local jumps on each two-state subsystems with rates in Eq.~\eqref{eq:rates}, the interaction with the tape acts only within the degenerate (single-excitation) subspace of the machine $\{\ket{01}_{\rm m},\ket{10}_{\rm m}\}$ 
producing effective transitions between these two states given by rates.
\begin{equation} 
\label{tape_rates}
\Gamma_{\rm tape}^{\uparrow}=r\phi^{2}p_{0}^{\rm tape},
\qquad
\Gamma_{\rm tape}^{\downarrow}=r\phi^{2}p_{1}^{\rm tape},
\end{equation}
where $r$ is the collision rate and $\phi \ll 1$ quantifies the interaction strength between the tape bits and the machine. These jumps are associated to bit flips in the tape bits (see Fig.~\ref{fig:ModelD}), which motivates the definition of an effective (inverse-temperature-like) bias for the information reservoir (i.e. the tape) by means of the detailed balance relation, that is, 
\begin{equation} \label{tape_bias}
\delta_{\rm tape} \equiv \ln\!\left(\frac{\Gamma_{\rm tape}^{\uparrow}}{\Gamma_{\rm tape}^{\downarrow}}\right) = \ln\!\left(\frac{p_0^{\rm tape}}{p_1^{\rm tape}}\right).
\end{equation}
This bias quantifies the tendency of the tape to generate jumps $\ket{01}_{\rm m} \rightarrow \ket{10}_{\rm m}$ ($\delta_{\rm tape} \geq 0$) over the opposite ones, $\ket{10}_{\rm m} \rightarrow \ket{01}_{\rm m}$ in the machine ($\delta_{\rm tape} \leq 0$), as given by the abundance of $0$'s and $1's$, respectively, in the initial state of the tape (i.e. before starting to interact with the machine).

The main feature of this model allowing the informational operation of the machine can be intuitively understood as a competition between the bias induced by the tape bits distribution, $\delta_{\rm tape}$ in Eq.~\eqref{tape_bias}, and the bias induced by the thermal baths on the machine. Indeed, in the absence of the tape, the mere (local) contact of the machine subsystems with their respective baths induces a bias within the degenerate subspace of the machine given by
\begin{equation}
\delta_{\rm th} \equiv \ln\!\left(\frac{p_{10}^{\mathrm{eq}}}{p_{01}^{\mathrm{eq}}}\right)= (\beta_{\rm c}-\beta_{\rm h}) \Delta E \geq 0,
\end{equation}
which follows from the explicit form of the equilibrium distribution of the machine, $\bm{p}_{\rm m}^{\rm eq}$, in close analogy with the virtual-temperature picture of autonomous thermal machines~\cite{Brunner2012,Skrzypczyk2015,Silva2016,Erker2017}.
As a consequence, when the values of $\delta_{\rm tape}$ and $\delta_{\rm th}$ are different, the heat fluxes to the hot and cold reservoirs couple with the bit flip rate in the tape, as we will shortly see. On the contrary, when the tape bias in Eq.~\eqref{tape_bias} matches the intrinsic (equilibrium) bias of the machine, $\delta_{\rm tape}= \delta_{\rm th}$, the machine will relax to its equilibrium state $\bm{p}_{\rm m}^{\rm eq}$ in the long time run.
In this situation the probabilistic interaction between the machine and the tape leaves both of them statistically unchanged and no net fluxes can be generated.
This equilibrium condition corresponds, for fixed bath temperatures and fixed energetic gap $\Delta E$, to an initial distribution of the tape bits given by $\bm{p}_{\rm tape}^{\rm eq} = (\tilde{p}_0^{\rm tape}, 1 -\tilde{p}_0^{\rm tape})^T$, with
\begin{equation}
\label{eq:p1v}
\tilde{p}_0^{\rm tape} = \frac{1}{1+e^{-\delta_{\rm th}}}.
\end{equation}
Otherwise, if the bits in the tape are prepared according to any other distribution, $\bm{p}_{\rm tape} \neq \bm{p}_{\rm tape}^{\rm eq}$, the dynamical evolution leads to a non-trivial non-equilibrium steady state (NESS) with persistent fluxes.

\section{NESS thermodynamic currents}
\label{NESS_thermo}
We are interested in the long-time dynamics of the model, where the probability vector becomes time-independent and satisfies $\mathbb{W}\,\mathbf{p}_{\rm m}^{\mathrm{ss}}=0$. In this regime, the thermal and informational exchanges can be characterized by the following average currents on the machine
\begin{equation}
\begin{split}
    J_{\rm c} &= \Gamma_{\rm c}^{\uparrow} (p_{00}^{\mathrm{ss}}+p_{10}^{\mathrm{ss}}) - \Gamma_{\rm c}^{\downarrow}(p_{11}^{\mathrm{ss}}+p_{01}^{\mathrm{ss}}), \\
    J_{\rm h} &= \Gamma_{\rm h}^{\uparrow} (p_{00}^{\mathrm{ss}}+p_{01}^{\mathrm{ss}}) - \Gamma_{\rm h}^{\downarrow}(p_{11}^{\mathrm{ss}}+p_{10}^{\mathrm{ss}}), \\
    J_{\rm tape} &= \Gamma_{\rm tape}^{\downarrow}p_{10}^{\mathrm{ss}} - \Gamma_{\rm tape}^{\uparrow}p_{01}^{\mathrm{ss}},
\end{split}
\end{equation}
where all probabilities are evaluated with the stationary distribution $\mathbf{p}_{\rm m}^{\mathrm{ss}}$. The currents $J_c$ and $J_h$ provide the probability fluxes (net number of jumps per unit time) induced by the cold and hot reservoirs, respectively, while $J_{\rm tape}$ measures the tape-induced probability flux within the degenerate subspace of the machine. Stationarity implies that these currents are not independent, but they satisfy, $J_{\rm h} = -J_{\rm c} = J_{\rm tape}$. 

Since each thermal transition changes the machine energy by $\Delta E$, the corresponding average heat currents can be calculated as
\begin{equation} \label{eq:heat_currents}
\dot Q_{\rm c} = \Delta E \, J_{\rm c},
\qquad
\dot Q_{\rm h} = \Delta E \, J_{\rm h},
\end{equation}
with the convention that $\dot Q_\alpha>0$ denotes heat flowing from reservoir $\alpha = c, h$ into the machine. These two heat currents verify the first law of thermodynamics, namely $\dot{Q}_c + \dot{Q}_h = 0$ in the steady state. By contrast, the tape-induced transitions act only by inducing a bias on the degenerate states $\ket{01}_{\rm m}$ and $\ket{10}_{\rm m}$ and therefore carry no energy. The role of the tape is therefore purely informational, but  it controls the direction of the heat currents between the two thermal reservoirs. 
On the other side, the heat fluxes from the baths also modify the initial distribution of bits inside the tape. In the stationary regime, the outgoing distribution in the tape after interaction with the machine is given by $\bm{p}_{\rm tape}^\prime = (p_0'^{\, \rm tape}, p_1'^{\, \rm tape})^T$ with probability $p_0'^{\, \rm tape} = p_0^{\rm tape} + J_{\rm tape}/r$. Therefore, although the tape remains energetically neutral, it is generally transformed by the interaction which can either homogenize the outgoing tape distribution or mix it. This transformation is quantified by the rate of change in the Shannon entropy of the tape:
\begin{equation} \label{eq:Stape}
\begin{split}
\dot S_{\rm tape} &= r\bigl[H(\bm{p}_{\rm tape}')-H(\bm{p}_{\rm tape})\bigr] \\
&= -\delta_{\rm tape}J_{\rm tape}
-r D\left(\bm p'_{\rm tape}\Vert\bm p_{\rm tape}\right).
\end{split}
\end{equation}
where $H(\bm{p})=- \sum_i p_i \ln p_i$ denotes the Shannon entropy of the distribution $\bm{p}$ with elements $p_i$, and $D({\bm p} || {\bm q})$ in the second line denotes the Kullback-Leibler divergence between probability distributions ${\bm p}$ and ${\bm q}$. The Kullback-Leibler divergence is non-negative and becomes zero if and only if ${\bm p} = {\bm q}$.

A positive value of the rate $\dot S_{\rm tape} >0$ indicates that, as a consequence of the interaction with the machine, the output tape distribution of bits becomes more mixed, that is, the number of $0$'s and $1$'s are more similar to each other. On the other hand, $\dot S_{\rm tape}<0$ corresponds to a reduction of the tape informational entropy, making the tape more homogeneous (e.g. more $0$'s than $1$'s) than it was before interaction. We refer to this entropy-reduction operation as erasing information in the tape, since input bits in either $0$ or $1$ are (probabilistically) reset to $0$, in analogy to the nomenclature given in Landauer's erasure~\cite{Landauer1961, Parrondo2015}. 
The second line in Eq.~\eqref{eq:Stape} 
is derived in appendix~\ref{app:tape_entropy} and shows that the change in the entropy of the tape is controlled by the sign of the flux $\delta_{\rm tape} J_{\rm tape}$ composed by the probability current and bias on the tape (both of which may be independently positive or negative); it competes with the non-negative term $r D\left(\bm p'_{\rm tape}\Vert\bm p_{\rm tape}\right)\geq 0$, which becomes zero only when the operation of the machine leaves the distribution of the invariant, that is, whenever ${\bm p}_{\rm tape} = {\bm p}_{\rm tape}^{\rm eq}$. 

The second law of thermodynamics in the setup can be obtained from the non-negativity of the entropy production rate~\cite{Landi_2021}, and establishes a balance relation between the entropy fluxes in the setup (see e.g. Refs.~\cite{Mandal2013,Barato_2014,Strasberg2017}). However, depending on the level of control assumed in the final state of the information reservoir, the entropy production may include or not an extra reset term~\cite{Manzano2018}, leading to different second-law-like inequalities. This distinction is particularly relevant here, and is directly related to the use given to the tape. In particular, when the state of the outgoing tape after interaction with the machine is retained as a valuable product of the process, we have:
\begin{equation}
    \dot S_{\rm tape}
    -\beta_{\rm c}\dot Q_{\rm c}
    -\beta_{\rm h}\dot Q_{\rm h}
    \geq 0.
    \label{eq:second_law_eraser}
\end{equation}
The above inequality is particularly important when information in the tape is erased (i.e. when a reduction in entropy of the tape is operated by the machine) since in this case the process can be regarded as a procedure to prepare a low-entropy tape, to be used afterwards. 

On the other hand, whenever the final state of the tape bits after interaction is intended to be inaccessible or lost, an extra source of entropy production arises due to the loss of that information, and the corresponding second-law inequality reads:
\begin{equation}
\begin{split}
    &\dot S_{\rm tape} + r D({\bm p_{\rm tape}'}|| {\bm p_{\rm tape}})
    -\beta_{\rm c}\dot Q_{\rm c}
    -\beta_{\rm h}\dot Q_{\rm h} \\ &= \, -\delta_{\rm tape} J_{\rm tape}  -\beta_{\rm c}\dot Q_{\rm c}
    -\beta_{\rm h}\dot Q_{\rm h}    \geq 0,
    \label{eq:second_law_fridge}
\end{split}
\end{equation}
where the extra term $r D({\bm p_{\rm tape}'}|| {\bm p_{\rm tape}})$ quantifies the entropic cost of resetting the outgoing tape bits back to their initial distribution. The above inequality is instead relevant in regimes where the information reservoir acts as an entropy sink, $\dot{S}_{\rm tape} > 0$, in which the tape distribution after interaction with the machine ${\bm p^\prime_{\rm tape}}$ is discarded, or physically reset back to its initial distribution ${\bm p_{\rm tape}}$.

Due to the stochastic nature of the interactions between the tape and the reservoirs, besides their average values, the steady state currents exhibit non-negligible fluctuations. These fluctuations become particularly relevant for small thermal machines where they affect the precision of the currents and hence of the thermodynamic operation performed by the machine~\cite{Campisi_2015,Pietzonka_2018,
Miller_2020,Saryal_2021}. The average values of the currents and their fluctuations can be computed using techniques from Full Counting Statistics~\cite{Esposito2009b} and Large Deviation Theory~\cite{Touchette2009}. 

Let $X_\alpha(t)$ denote the time-integrated probability current associated with channel
$\alpha\in\{{\rm c}, {\rm h}, {\rm tape}\}$. In the long-time limit, the average integrated currents verify 
\begin{equation} 
\label{eq:integrated}
\langle X_\alpha(t)\rangle \simeq tJ_\alpha,
\qquad
\operatorname{Var}[X_\alpha(t)]
\simeq t\,\operatorname{Var}(J_\alpha),
\end{equation}
where $\operatorname{Var}(J_\alpha)$ denotes the asymptotic variance rate of the probability current of $J_{\alpha}$ or current dispersion rate. In the present model, conservation of the two local occupations constrains the three integrated currents up to bounded boundary terms. As demonstrated in appendix~\ref{app:current_statistics}, this leads to the simplifying relation for the asymptotic variance rates: 
\begin{equation}
    \operatorname{Var}[\dot Q_{\rm c}]
    =
    \operatorname{Var}[\dot Q_{\rm h}]
    =
    \Delta E^2\,\operatorname{Var}[J_{\rm tape}],
\end{equation}
where ${\rm Var}[\dot Q_{\rm c}] = \Delta E^2 {\rm Var}[J_{\rm c}]$ and ${\rm Var}[\dot Q_{\rm h}] = \Delta E^2 {\rm Var}[J_{\rm h}]$ are the corresponding variances of the hot and cold heat currents. These variances rates, as well as all other cumulants are obtained from derivatives of the dominant eigenvalue of the tilted dynamical generator.

In this context, the thermodynamic uncertainty relation (TUR) relates the precision of the non-equilibrium steady state currents with the dissipation of the machine as~\cite{Barato2015,Gingrich2016, Horowitz2020}:
\begin{equation} \label{eq:TUR}
    \frac{\operatorname{Var}[\dot Q_\alpha]}
    {\dot Q_\alpha^2} 
    \geq \frac{2}{\dot S_{\rm tot}},
\end{equation}
with $\dot S_{\rm tot} = -\delta_{\rm tape} J_{\rm tape}  -\beta_{\rm c}\dot Q_{\rm c} -\beta_{\rm h}\dot Q_{\rm h}$ the entropy production rate in the case of discarding the output tape distribution, corresponding to Eq.~\eqref{eq:second_law_fridge}, and $\alpha = {\rm h, c}$. The TUR implies that increasing the precision of any of the thermodynamic currents incurs in additional dissipation, unveiling a fundamental trade-off between these two quantities. The TUR can be saturated at the equilibrium point or in the linear response regime, where it has been recently shown to corresponds to the Pareto-optimal solution among output power, dissipation, efficiency and fluctuations~\cite{almanza2026}.

We can rewrite the TUR in Eq.~\eqref{eq:TUR} by using the explicit expression of $\dot{S}_{\rm tot}$ and the relation between probability currents in the NESS as 
\begin{equation}
    \frac{|\dot Q_{\rm c}|}
    {\operatorname{Var}[\dot Q_{\rm c}]}
    \leq
    \frac{|\delta_{\rm tape}-\delta_{\rm th}|}{2\Delta E},
    \label{eq:tur_ratio_heat}
\end{equation}
where we took the particular case of $\alpha= {\rm c}$ in Eq.~\eqref{eq:TUR}. The above equation reveals that the signal-to-noise ratio (inverse Fano factor) $|\dot Q_{\rm c}|/{\rm Var}[\dot Q_{\rm c}]$ of the cold heat current cannot be increased indefinitely, but it is upper bounded by the differences in the tape and equilibrium biases introduced above, pinpointing $\delta_{\rm tape} - \delta_{\rm th}$ as a relevant thermodynamic resource in the setup (a similar relation is obtained for the hot heat current, see App.~\ref{app:current_statistics}).

\begin{figure}
    \centering
    \includegraphics[width=\linewidth]{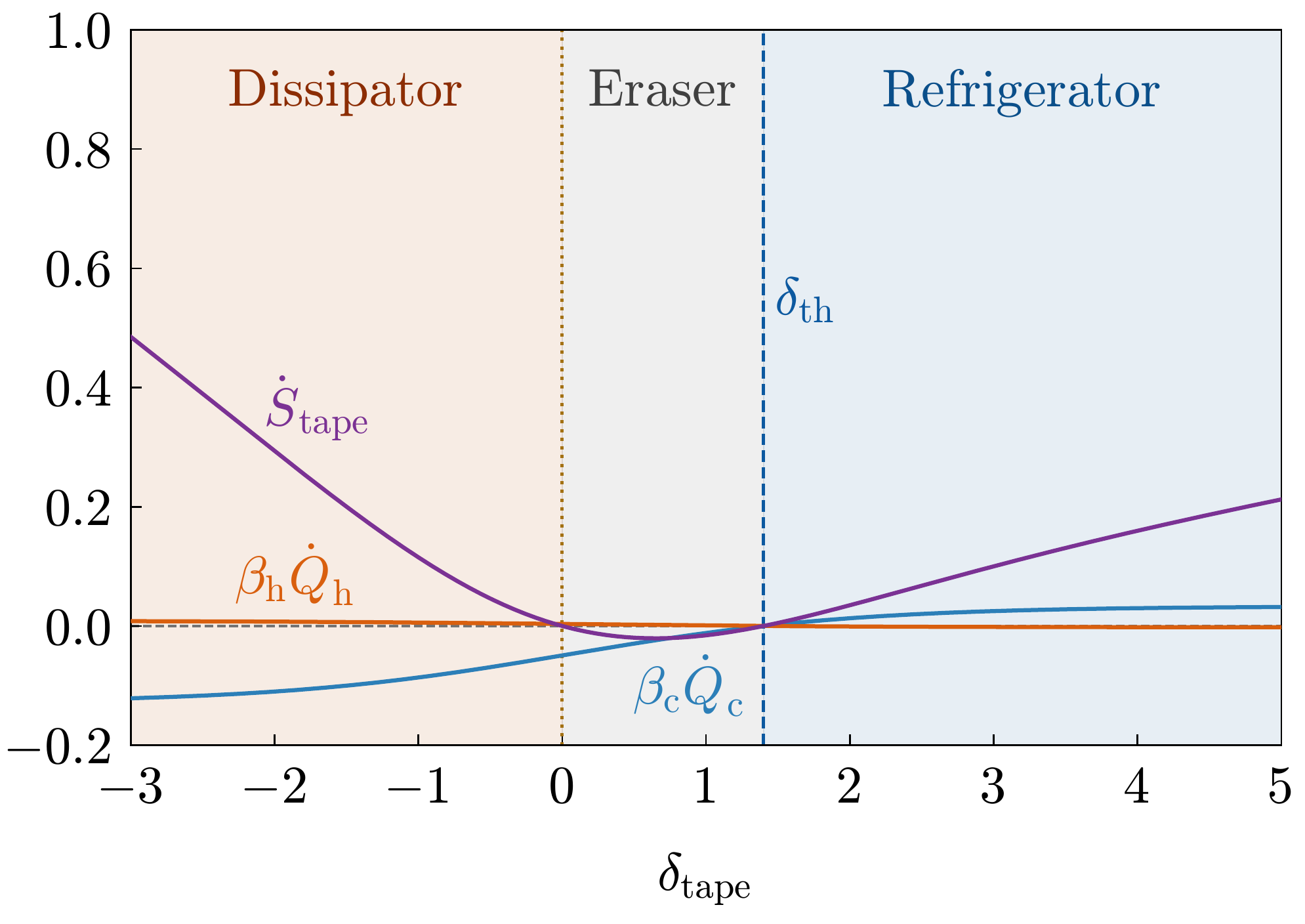}
    \caption{
    Main currents in the NESS along different operating regimes as a function of the tape bias $\delta_{\rm tape}$, for fixed bath temperatures and energy spacing $\Delta E$. The blue and orange curves show the (scaled) heat currents from the cold
    $\beta_{\rm c} \dot Q_{\rm c}$ and hot $\beta_{\rm h} \dot Q_{\rm h}$ baths, while the purple curve corresponds to the tape entropy change rate, $\dot S_{\mathrm{tape}}$. The vertical dotted line at $\delta_{\rm tape}=0$
    marks the unbiased tape ($p_0^{\rm tape} = 1/2$), whereas the dashed line represents the thermal gradient $\delta_{\rm th}$ ($p_0^{\rm tape} = \tilde p^{\rm tape}_0$) where all stationary currents vanish. The shaded colors identify the different modes od operation of the machine. All currents are given in units of $r$. }
    \label{fig:currents}
\end{figure}

\section{Operational regimes}
\label{sec:thermodynamic_regimes}

We find three different regimes of operation that the informational machine can perform when running in non-equilibrium steady state conditions, as characterized by the sign of the heat currents and informational fluxes:
\begin{equation}
\begin{array}{lll}
\dot{S}_{\rm tape} < 0,\ \, \dot Q_{\rm c}<0,\ \,\dot Q_{\rm h}>0,
& \text{Eraser},
\\[2mm]
\dot{S}_{\rm tape} > 0,\ \, \dot Q_{\rm c}>0,\ \,\dot Q_{\rm h}<0,
& \text{Refrigerator}, 
\\[2mm]
\dot{S}_{\rm tape} > 0,\ \, \dot Q_{\rm c}<0,\ \,\dot Q_{\rm h}>0,
& \text{Dissipator}.
\end{array}
\label{eq:three_regimes}
\end{equation}

For fixed bath temperatures $\beta_{\rm c} \geq \beta_{\rm h}$ and energy gap $\Delta E$, the operating mode of the machine is controlled by the distribution of the incoming tape bits, $\bm{p}_{\rm tape}$, determining the bias $\delta_{\rm tape}$ in Eq.~\eqref{tape_rates}, as illustrated in Fig.~\ref{fig:currents}. 
In the eraser regime (grey central area in Fig~\ref{fig:currents}), the spontaneous heat flux from the hot reservoir $\dot{Q}_{\rm h} >0$ to the cold one $\dot{Q}_{\rm c} < 0$ is used by the machine to generate a reduction of entropy in the tape $\dot{S}_{\rm tape} < 0$. This reduction of entropy in the tape is interpreted as erasure of information in the sense of Landauer's principle~\cite{Parrondo2015}, that is, whatever the initial values of the bits are, the number of bits in state $0$ after interaction with the machine will increase. This regime requires a tape bias satisfying $0<\delta_{\rm tape}<\delta_{\rm th}$ (that is, $1/2 < p_0^{\rm tape} < \tilde{p}_0^{\rm tape}$). In this way, the output probability of the tape bits verify  $p_{0}^{\prime ~\rm tape} > p{_0}^{~\rm tape} $. As the tape bias increases to approach the equilibrium value, all the fluxes decrease until they vanish at $\delta_{\rm tape} = \delta_{\rm th}$, where the tape distribution is left unchanged $p_{0}'^{~\rm tape} = p_0^{\rm tape}$.

Moreover, when the tape bias is strong enough, $0< \delta_{\rm th}< \delta_{\rm tape}$, the spontaneous heat flow can be reversed, and the machine absorbs heat from the cold bath $\dot{Q}_{\rm c} > 0$ and releases it into the hot one, $\dot{Q}_{\rm h} < 0$. Such a refrigeration of the cold bath can only be accomplished at the cost of the increase in entropy of the tape, $\dot{S}_{\rm tape} > 0$, whose output distribution becomes now more mixed as a consequence of the interaction with the machine ($p^{\prime \, \rm tape}_0 < p_{0}^{~\rm tape}$ ). This leads to an information-assisted refrigerator regime (right blue region in Fig.~\ref{fig:currents}), which typically requires an initial distribution of the tape bits quite ordered, i.e. $1/2 < \tilde{p}_0^{\rm tape}< {p}_0^{\rm tape}$.

Finally, in the dissipator regime (left orange area in Fig.~\ref{fig:currents}) the flux of heat from the hot bath $\dot{Q}_{\rm h} > 0$, to the cold one $\dot{Q}_{\rm c}<0$, is accompanied by an increase in the tape entropy $\dot{S}_{\rm tape} > 0$, i.e. all currents follow the spontaneous directions set by their natural biases. This regime is obtained from the eraser regime when the bias in the tape is inverted, $\delta_{\rm tape} < 0$, or equivalently when $p_0^{\rm tape} < 1/2$. That is, when the tape is initially populated with more bits in state $1$ than in state $0$, the increase in the number of zeros, $p_{0}'^{~\rm tape} > p_0^{\rm tape}$, operated by the machine using the spontaneous flow of heat from hot to cold baths tends now to increase its entropy. 

\begin{figure}[t]
    \centering
    \includegraphics[width=0.49\columnwidth]{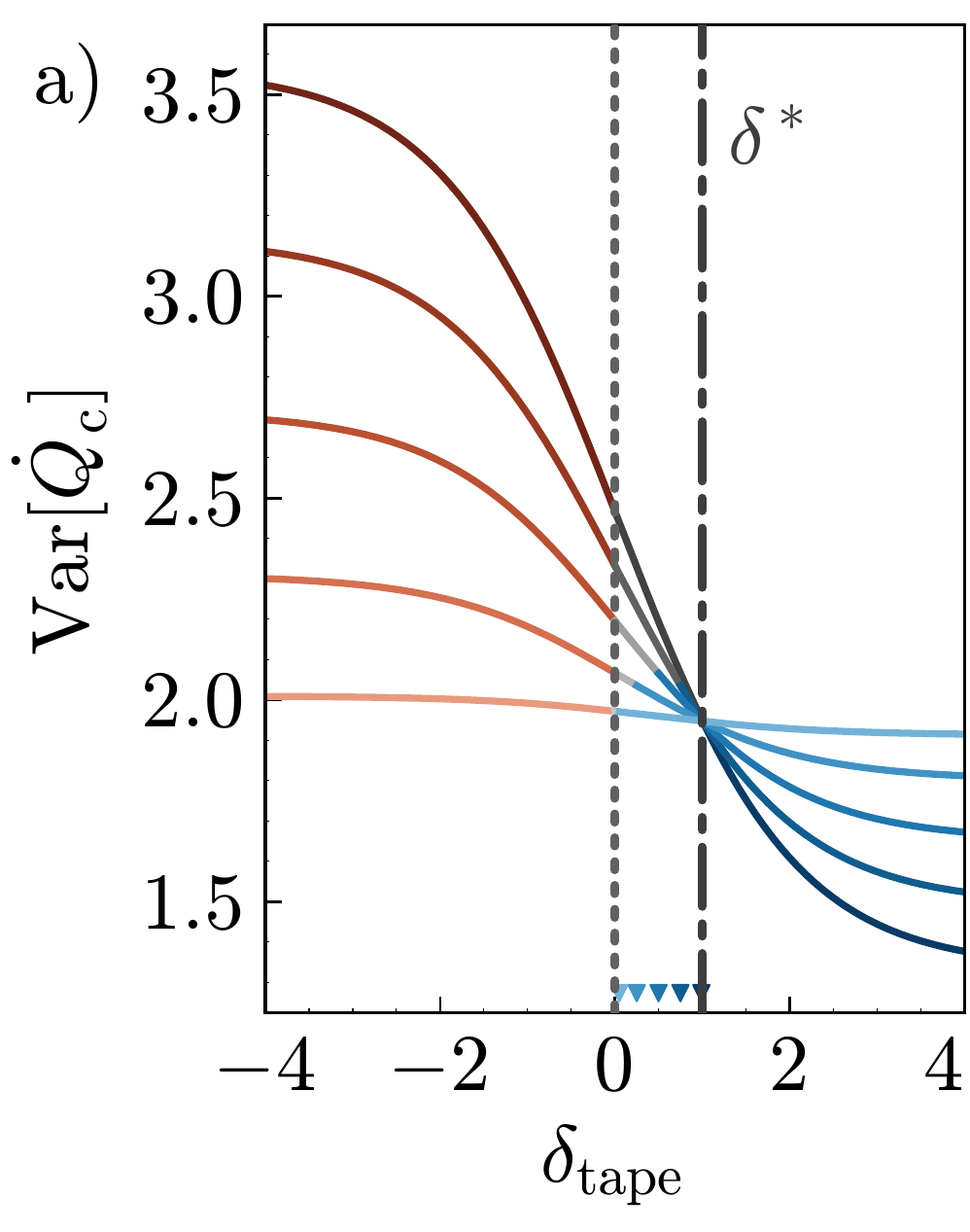}
    \hfill
    \includegraphics[width=0.49\columnwidth]{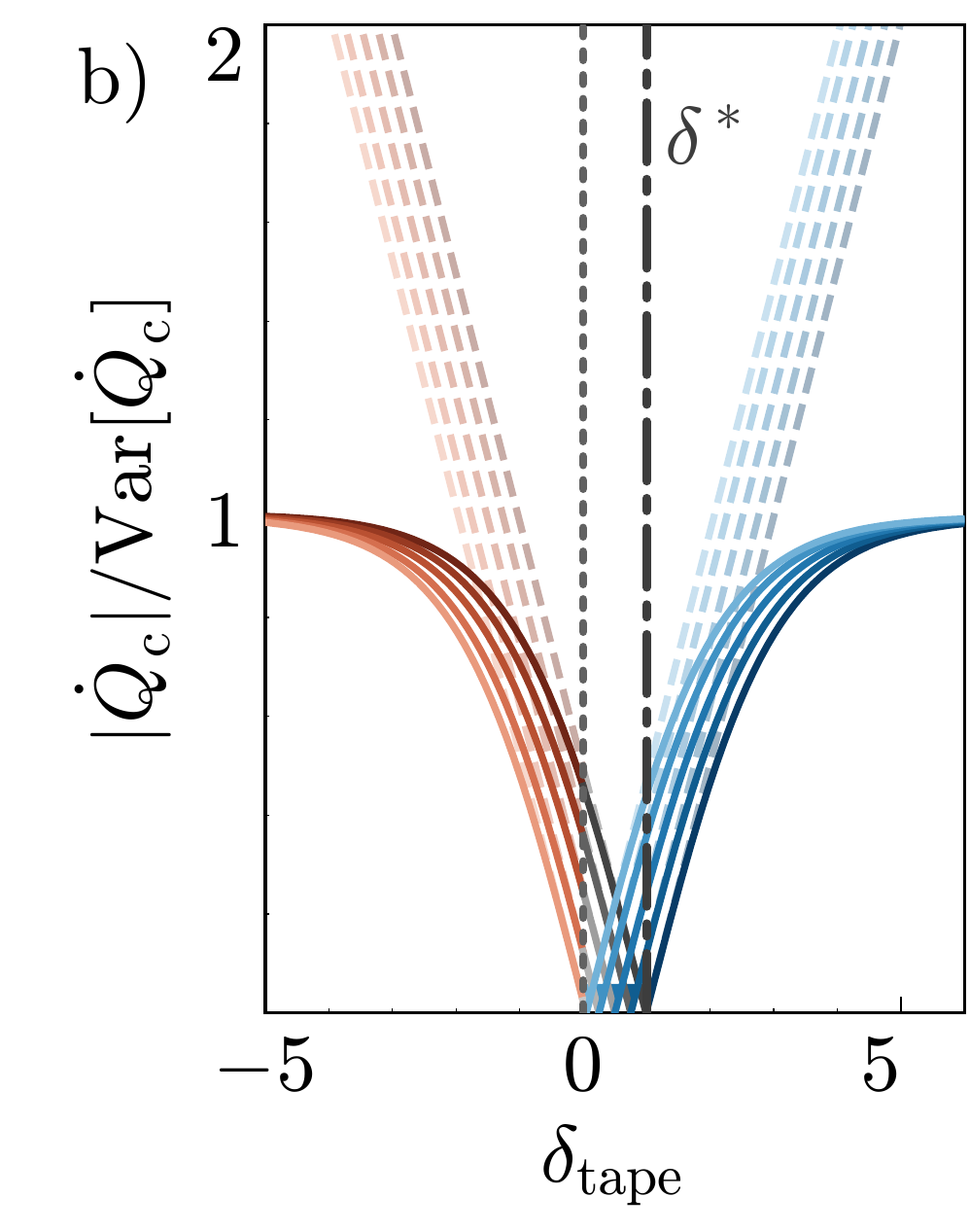}
    \caption{
    Current fluctuations and thermodynamic precision of the informational
    machine.
    (a) Asymptotic variance of the cold heat current as a function of the incoming tape population bias. The different curves correspond, from darker to lighter colors to values of $\beta_{\rm h}/\beta_{\rm c}= \{ 0.01,\,0.25,\,0.50,\,0.75, 0.95\}$ and the color tone represents the operational regimes, being the red tone for Dissipator, grey for Eraser and blue for refrigerator. The dotted vertical line marks the unbiased tape, $\delta_{\rm tape}=0$, and the
    dash-dotted line indicates the characteristic crossing point $\delta_\ast$. The triangular markers denote the equilibrium points $\delta_{\rm th}$ of the different curves. (b) Signal-to-noise ratio of the cold current for same values of $\beta_{\rm h}/\beta_{\rm c}$ and color ordering (solid curves) and corresponding TUR bounds in Eq.~\eqref{eq:tur_ratio_heat} (dashed curves).}
    \label{fig:current_fluctuations}
\end{figure}

The behavior of the fluctuations in the currents across the three operational regimes of the machine is illustrated in Fig.~\ref{fig:current_fluctuations} through the variance of the heat current from the cold bath and its precision. 

As can be appreciated in Fig.~\ref{fig:current_fluctuations}(a), the fluctuations in the heat current from the cold bath are progressively reduced when increasing the density of zeros in the incoming tape (increasing $\delta_{\rm tape}$), specially for high values of the temperature gradient, $\beta_{\rm c}/\beta_{\rm h}$. We recall that, contrary to the average currents, the variance do not vanish at the equilibrium point $\delta_{\rm tape} = \delta_{\rm th}$
but fluctuations remain active. In addition, all curves display a crossing point [dashed vertical line in Fig.~\ref{fig:current_fluctuations}(a)] approximately at:
\begin{equation}
    \delta_{*} \simeq \beta_{\rm c} \Delta E,
    \label{eq:p0star_main}
\end{equation}
which is always placed inside the refrigerator regime,  i.e. $\delta_\ast \geq \delta_{\rm th}$. The above expression is derived in appendix~\ref{app:fluctuation_crossing}. It corresponds to the point in which the fluctuations become independent of the temperature gradient, to leading order in the tape-machine effective coupling $(r\phi^2 \ll 1)$ corresponding to slow collisions. This may suggest that arbitrary high precision in the currents could be reached at this point by increasing the temperature gradient. However, such an enhancement is spoiled by the fact that  
the equilibrium point (where all average currents vanish) shifts towards the crossing point, i.e. $\delta_{\rm th} \rightarrow \delta_{\ast}$ when $\beta_{\rm h}/\beta_{\rm c} \rightarrow 0$ [see colored triangles in the bottom of Fig.~\ref{fig:current_fluctuations}(a)].

The tradeoff between precision and dissipation imposed by the TUR in Eq.~\eqref{eq:tur_ratio_heat} is shown in Fig.~\ref{fig:current_fluctuations}(b). There the signal-to-noise ratio of the cold heat current $|\dot Q_{\rm c}|/{\operatorname{Var}[\dot Q_{\rm c}]}$ is displayed as a function of $p_0^{\rm tape}$ for different values of the temperature gradient (solid lines), and it is compared to the TUR bound in Eq.~\eqref{eq:tur_ratio_heat}, $|\delta_{\rm tape}-\delta_{\rm th}|/(2\Delta E)$ (dashed lines). The small values of the signal-to-noise ratio highlight the noisy nature of the currents. The ratio approaches the bound close to the equilibrium point ($\delta_{\rm tape} \rightarrow \delta_{\rm th}$) where the machine enters the linear-response regime, and is saturated exactly on it, when $|\dot{Q}_{\rm c}| \rightarrow 0$. Farther from that point the TUR becomes progressively less tight, and the curves for the signal-to-noise ratio converge to a maximum value of $1$ at both edges of the plot. In these limits the current fluctuations approach an independent Poisson process for which ${\rm Var}[{\dot{Q}_{\rm c}}] \rightarrow |\dot{Q}_{\rm c}|$. 

\section{Informational efficiencies and their fluctuations}
\label{sec:efficiencies}
Having characterized the stationary operating regimes, we now examine the performance of the two thermodynamically useful modes of operation, erasure and refrigeration, in terms of their efficiency. When moving apart from typical thermal configurations, the notion of efficiency itself may becomes more intricate~\cite{hybrid2020}. For that, we introduce suitable definitions of efficiency in each regime and explore both their average (long-time) values and their fluctuations, extending the theory of stochastic efficiency fluctuations to this informational engine~\cite{Verley2014a,Verley2014b,Manikandan2019}.

\subsection{Heat-driven eraser efficiency}
\label{subsec:eraser}

As mentioned before, in the eraser regime of operation, we identify the main useful output as the rate at which entropy is
removed from the tape, $-\dot{S}_{\rm tape}$, while the main input resource is given by the heat current from the hot bath, $\dot{Q_{\rm h}}$. As a consequence the (macroscopic) thermodynamic efficiency in this regime is defined as:
\begin{equation}
    \bar{\eta}
    \equiv
    \frac{- \dot S_{\rm tape}}
         {\beta_{\rm c} \dot Q_{\rm h}}
    \leq
    \eta_{\rm C},
    \label{eq:eraser_efficiency}
\end{equation}
where $\beta_{\rm c}$ acts as a reference (inverse) temperature to transform entropic quantities into energetic ones~\footnote{That is, the useful output in the numerator of $\bar{\eta}$ expressed in energy units reads $-k_B T_{\rm c} \dot{S}_{\rm tape}$}. The choice of the cold temperature as the reference stands in analogy with traditional heat engines, as elaborated in Ref.~\cite{hybrid2020}. Here above $\eta_{\rm C} \equiv 1-\beta_{\rm h}/\beta_{\rm c}$ is Carnot's efficiency, and the inequality is obtained from the second-law inequality in Eq.~\eqref{eq:second_law_eraser} using $\dot Q_{\rm h}=-\dot Q_{\rm c}$. Since $\dot S_{\rm tape}$ is negative during erasure, and $\dot{Q}_{\rm h}$ is positive, the efficiency $\bar{\eta}$ is non-negative and bounded by $\eta_{\rm C}$, similarly to work producing thermal machines, i.e. $\bar{\eta}/\eta_{\rm C}\leq 1$. The bound is approached close to the equilibrium point (i.e. $p_0^{\rm tape}\rightarrow\tilde p_0^{\, \rm tape}$) where the thermal and informational biases compensate each other $\delta_{\rm tape} \rightarrow \delta_{\rm th}$, and all currents vanish. Reversible erasure (i.e. Landauer's limit) is thus reached only in the zero-output limit $\dot{S}_{\rm tape} \rightarrow 0$.

Similarly to the currents, fluctuations may affect the values of efficiency, which can be also defined at the stochastic level~\cite{Verley2014a,Verley2014b}. An stochastic efficiency can then be defined from the ratio of the relevant time-integrated probability currents $X_{\alpha}(t)$ with $\alpha=\{\rm c, h, tape \}$ [see Eq.~\eqref{eq:integrated}]. For the eraser regime, the input resource corresponds to the time-integrated heat current from the hot reservoir, $q_{\rm h}(t) \equiv \Delta E X_h(t)$. However, the relevant output is not exactly identified with the integrated probability current $X_{\rm tape}(t)$ associated to $J_{\rm tape}$, but with the stochastic changes in the tape self-information associated to $\dot{S}_{\rm tape}$ [c.f. Eq.~\eqref{eq:Stape}]. Therefore we define: 
\begin{equation}
    \eta(t)
    =
    -\frac{\Delta s_{\rm tape}(t)}
    {\beta_{\rm c} \, q_{\rm h}(t)},
    \label{eq:stochastic_eraser_eff}
\end{equation}
with $\Delta s_{\rm tape}(t)$ the accumulated changes over the system-tape interactions in the tape surprisal (see appendix~\ref{app:efficiency_fluctuations}). This quantity should be distinguished from the macroscopic (average) efficiency $\bar{\eta}$ in Eq.~\eqref{eq:eraser_efficiency}, defined directly from the stationary average currents. 
\begin{figure}[tbp]
    \centering
    \includegraphics[width=\columnwidth]{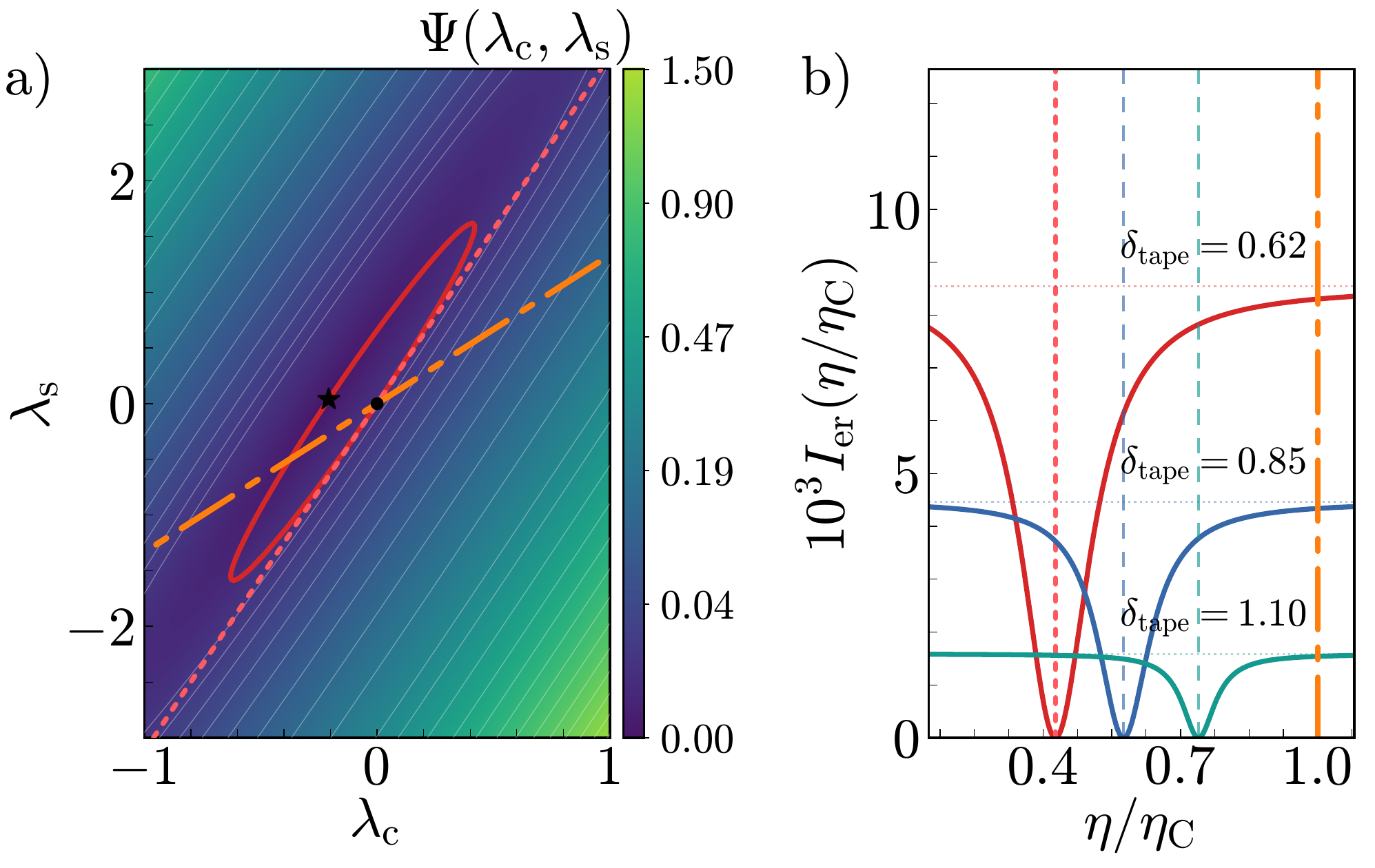}
    \caption{
    Stochastic efficiency fluctuations of the information eraser.
    (a) Shifted joint scaled cumulant generating function
    $\Psi(\lambda_c,\lambda_s)$ for $p_0^{\rm tape}=0.65$.
    The solid red curve represents the locus of constrained minima
    $\tilde{\boldsymbol{\lambda}}(\eta/\eta_{\rm C})$ obtained by varying the reduced
    efficiency.
    The pink dashed and orange dash-dotted lines indicate the
    macroscopic and reversible efficiencies, respectively.
    The black circle marks the origin, while the black star
    identifies the global minimum of the SCGF, associated with
    the least-probable efficiency.
    (b) Efficiency rate functions $I_{\rm er}(\eta/\eta_{\rm C})$
    for $\delta_{\rm tape}=0.62$, $0.85$, and $1.10$,
    corresponding to $p_0^{\rm tape}=0.65$, $0.70$, and $0.75$,
    respectively.
    The colored dashed vertical lines indicate the macroscopic
    efficiencies, where the rate functions vanish, while the
    orange dash-dotted line marks the reversible efficiency
    $\eta/\eta_{\rm C}=1$.
    The horizontal dotted lines indicate the corresponding
    asymptotic plateau values.}
    \label{fig:eraser_reversible}
\end{figure}

The fluctuations of $\eta(t)$ in the long-time limit can be assessed using Large Deviation Theory~\cite{Touchette2009}. The joint fluctuations of $\Delta s_{\rm tape}(t)$ and $q_{\rm h}(t)$, are encoded in the Scaled Cumulant Generating Function (SCGF):
\begin{equation}
\Psi(\lambda_{\rm h},\lambda_{\rm s})=\lim_{t\rightarrow\infty}\frac{1}{t}\ln\left\langle e^{\lambda_{\rm h} \beta_{\rm c} q_{\rm h}(t)+\lambda_{\rm s} \Delta s_{\rm tape}(t)} \right\rangle,
\label{eq:eff_scgf}
\end{equation}
where $\lambda_{\rm h}$ and $\lambda_s$ are the associated counting fields~\cite{Esposito2009b}. The derivatives of the SCGF evaluated at $\lambda_i = 0$  for $i={\rm h, s}$, determine the average currents $\dot{Q}_{\rm h}$ and $\dot{S}_{\rm tape}$, respectively, together with their covariance matrix. The distribution of the (normalized) stochastic efficiency hence satisfies, in the long-time limit, the large-deviation principle~\cite{Verley2014a,Verley2014b}: 
\begin{equation}
P_\tau \left(\frac{\eta}{\eta_{\rm C}}\right)\asymp e^{-\tau I_{\rm er}\left(\frac{\eta}{\eta_{\rm C}}\right)},
\label{eq:eraser_eff_ldf}
\end{equation}
where $I_{\rm er} (\eta/\eta_{\rm C}) := -\min_{\lambda} \Psi(\eta\lambda,\lambda)$ is the corresponding efficiency rate function, obtained from the contraction of the SCGF in Eq.~\eqref{eq:eff_scgf}.

The above construction for assessing the long-time fluctuations of the stochastic efficiency has a geometrical interpretation elaborated in Refs.~\cite{Verley2014a,Verley2014b,Manikandan2019}, which we illustrate in Fig.~\ref{fig:eraser_reversible} for our informational machine for a fixed value of $\delta_{\rm tape}$. For each fixed value of the stochastic efficiency $\eta/\eta_{\rm C}$, we minimize the SCGF along the straight line $\lambda_{\rm c}=\eta \, \lambda_{\rm s}$. The locus of the resulting minima defines the red curve shown in Fig.~\ref{fig:eraser_reversible}(a), from which the corresponding efficiency rate function can be obtained (see red curve in panel Fig.~\ref{fig:eraser_reversible}(b) corresponding to $\delta_{\rm tape} = 0.62$). The geometrical interpretation of Fig.~\ref{fig:eraser_reversible}(a) directly identifies two characteristic efficiencies. The macroscopic efficiency in Eq.~\eqref{eq:eraser_efficiency} corresponds to the dotted line tangent to the red circle which crosses the origin (black circle). This leads to a minimum of the rate function $I_{\rm er}(\bar{\eta}/\eta_{\rm C})=0$, hence identifying the value $\bar{\eta}$ as the most probable efficiency for long times. Conversely, the largest value of $I_{\rm er}(\eta/\eta_{\rm C})$ is reached by an efficiency line $\lambda_{\rm c}=\eta_* \, \lambda_{\rm s}$ that intersects the global minimum of the SCGF (black star), which corresponds to the least-probable efficiency. For conventional stochastic heat engines satisfying the appropriate fluctuation symmetry, this least-probable value coincides with the Carnot efficiency~\cite{Verley2014a,Verley2014b}.  However, this identification can be lost when the symmetry conditions required by the standard construction are not fulfilled~\cite{Gingrich2014,Manikandan2019}. This is precisely the situation for the present eraser. As shown in Fig.~\ref{fig:eraser_reversible}(a), the Carnot efficiency line $\lambda_{\rm s} = \eta_C \lambda_{\rm h}$ (dashed-dotted line) does not pass through the global minimum of the SCGF (black star), and therefore does not coincide with the least-probable efficiency.

The explicit form of the efficiency rate functions $I_{\rm er}(\eta/\eta_{\rm C})$ are shown in Fig.~\ref{fig:eraser_reversible}(b) for different values of the tape bias. Each curve vanishes at the corresponding macroscopic efficiency (vertical dashed lines) and increases rapidly away from this most typical value, approaching a broad plateau close to the maximum value. In particular, Carnot's efficiency (vertical dashed-dotted line at $\eta/\eta_{\rm C}=1$) lies within this broad region but is not singled out as a unique maximum of $I_{\rm er}$. As the incoming tape approaches the equilibrium point ($\delta_{\rm tape} \rightarrow \delta_{\rm th}$), the macroscopic efficiency moves towards $\eta_C$ while the overall height of the plateau decreases. Reversible erasure therefore becomes typical only in the same limit in which the stationary erasure current vanishes.

The absence of a local maximum in $I_{\rm er}(\eta/\eta_{\rm C})$ matching Carnot's efficiency originates from the nature of the useful output current in the eraser regime, $\dot{S}_{\rm tape}$. Unlike an integrated thermodynamic current $X_{\alpha}(t)$, the stochastic entropy change $\Delta s_{\rm tape}(t)$ is not generally antisymmetric under time reversal (see App.~\ref{app:efficiency_fluctuations}). Consequently, the joint SCGF in Eq.~\eqref{eq:eff_scgf} does not satisfy the fluctuation symmetry that, in the standard theory of efficiency fluctuations, singles out Carnot's efficiency as the least-probable one~\cite{Verley2014a,Verley2014b,Manikandan2019}.

\subsection{Information-powered fridge efficiency}
\label{subsec:refrigerator}

In the refrigeration regime, the role of the tape is qualitatively different. As commented before, the incoming tape in this case acts as an entropy sink, where its distribution (quantified by $\delta_{\rm tape}$) provides an informational resource that bias transitions in the thermal machine, and is able to drive heat from the cold bath to the hot one (when $\delta_{\rm tape}> \delta_{\rm th}$). The main useful output is therefore $\dot{Q}_{\rm c}$. However, in this case the input informational resource driving refrigeration is not $\dot{S}_{\rm tape}$, but $\delta_{\rm tape} J_{\rm tape}$, i.e. the actual current driving transitions (and heat currents) in the machine. This point can also be understood from the fact that the state of the tape after interaction (entering in $\dot{S}_{\rm tape}$) is not relevant for the cooling operation. This situation is similar to other standard schemes of information reservoirs~\cite{Strasberg2017}. 

Taking into account the relevant input resource and output current, we define the (macroscopic) coefficient of performance (COP):
\begin{equation}
    \bar{\epsilon}
    \equiv
    \frac{ \beta_{\rm h} \dot Q_{\rm c}}
         {-\delta_{\rm tape}J_{\rm tape}}
    \leq \bar{\epsilon}_{\rm C},
    \label{eq:refrigerator_cop}
\end{equation}
where here we used $\beta_{\rm h}$ as the reference (inverse) temperature, in analogy to power-driven refrigerators (see Ref.~\cite{hybrid2020} for details). The upper bound follows from the second law inequality in Eq.~\eqref{eq:second_law_fridge}, where $\bar{\epsilon}_{\rm C}= \beta_{\rm h}/(\beta_{\rm c}-\beta_{\rm h})$ denotes the maximum (reversible) COP.

The macroscopic COP in Eq.~\eqref{eq:refrigerator_cop} remains positive throughout the refrigerator regime, where $\dot Q_{\rm c}>0$ and $\delta_{\rm tape} \,J_{\rm tape}<0$.
Combining the relation $\dot Q_{\rm c}=-\Delta E J_{\rm tape}$ [c.f. Eq.~\eqref{eq:heat_currents}] with the inequality in Eq.~\eqref{eq:refrigerator_cop} we obtain the so-called cooling window~\cite{Alonso2013} of the information-powered fridge:
\begin{equation} \label{eq:cwindow}
    \frac{\bar{\epsilon}}{\epsilon_{\rm C}}
= \frac{\delta_{\rm th}}{\delta_{\rm tape}}\leq 1
\end{equation}
which imposes $\delta_{\rm tape} \geq \delta{\rm th}$ for the working regime of the fridge, in agreement with the analysis in Sec.~\ref{sec:thermodynamic_regimes}. As the tape bias approaches the thermal one, $\delta_{\rm tape}\to\delta_{\rm th}$, the COP approaches its maximum value while the currents vanish.

In analogy with the eraser regime, we now consider the joint fluctuations of the cold heat current and the input entropic flux leading to the macroscopic COP in Eq.~\eqref{eq:refrigerator_cop}. The corresponding stochastic COP can be defined as the ratio between the time-integrated heat current from the cold reservoir, $q_{\rm c}(t) = \Delta E~ X_{\rm c}(t)$, and the time-integrated probability current (scaled in suitable energy units) $\beta_{\rm h}^{-1} \delta_{\rm tape} X_{\rm tape}(t)$. That is:
\begin{equation}
\epsilon(t) = \frac{\beta_{\rm h} \, q_{\rm c}(t)}
{-\delta_{\rm tape} X_{\rm tape}(t)}.
\end{equation}
We are interested in the fluctuations of $\epsilon(t)$ in the long-time limit. In this case, at difference from the eraser regime, the relevant stationary currents and their fluctuations are not independent between them, as discussed in Sec.~\ref{NESS_thermo} and App.~\ref{app:current_statistics}. Indeed, the time-asymptotic accumulated currents entering the stochastic COP definition satisfy $X_{\rm tape}(t) \approx -X_{\rm c}(t)$, since the boundary terms vanish in the long time run. Consequently, for sufficiently long times, every realization approaches the same (macroscopic) COP, $\bar{\epsilon}$ in Eq.~\eqref{eq:cwindow}. The efficiency fluctuations therefore vanish, corresponding to the tight coupling limit of stochastic efficiency fluctuations~\cite{Verley2014b,Manikandan2019}. Similarly, the probability distribution for the stochastic COP, $P_\tau \left(\epsilon/\epsilon_C \right)$, becomes a delta function around the macroscopic COP in the long-time limit.  
Here efficiencies different from $\bar{\epsilon}$ form a degenerate least-probable sector, as it is characteristic of machines under tight coupling conditions~\cite{Manikandan2019}. Close to equilibrium, the stationary currents vanish, while $\bar{\epsilon}$ approaches the maximum (reversible) COP.

\section{Optimizing the performance}
\label{sec:Optimization}
With suitable definitions for power and efficiency in the two main operating regimes of the informational engine, we now focus in their optimization. In particular we determine the optimal power output in each regime across energy and temperature scales and quantify the associated trade-off with the respective thermodynamic efficiency.

\subsection{Optimal Erasure}

\begin{figure}[t]
    \centering
    \includegraphics[width=\columnwidth]{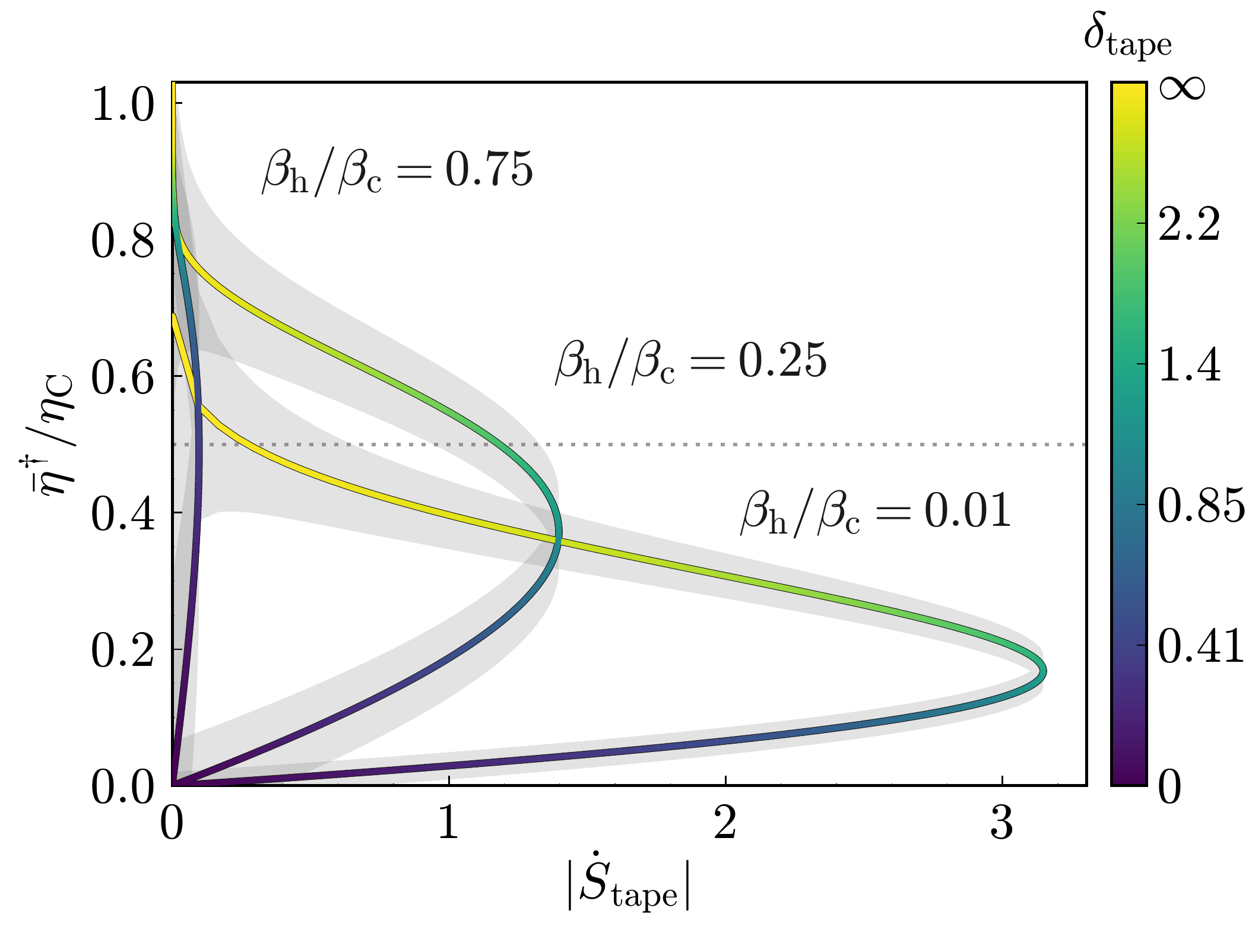}
    \caption{
Optimal performance of the information eraser. For each tape bias, $\delta_{\rm tape}$, the energetic scale $\beta_{\rm c}\Delta E$ is chosen to maximize the erasure rate $-\dot S_{\rm tape}$, and the reduced efficiency $\eta^\dagger/\eta_{\rm C}$ is evaluated at the same
operating point. The color scale indicates $\delta_{\rm tape}$, while the gray bands quantify the corresponding stochastic dispersion of the efficiency. Curves correspond to representative thermal ratios $\beta_{\rm h}/\beta_{\rm c}=0.01,\,0.25,$ and $0.75$.}
    \label{fig:eraser_performance}
\end{figure}

To characterize optimal erasure, we minimize the entropy change rate $\dot S_{\rm tape} < 0$ with respect to the energetic scale $\beta_{\rm c}\Delta E$ at fixed tape bias $\delta_{\rm tape}$. In the following, we denote by $\bar{\eta}^\dagger$ the macroscopic efficiency of the eraser in Eq.~\eqref{eq:eraser_efficiency} evaluated at the resulting optimal operating point. By repeating the optimization for different tape biases, we obtain power--efficiency curves which are shown in Fig.~\ref{fig:eraser_performance}, for three different values of the temperature ratio $\beta_{\rm h}/\beta_{\rm c}$. There the gray bands provide a notion of the efficiency fluctuations at maximum power, as quantified by the variance of the efficiency. The latter is calculated from the  local curvature of the efficiency rate function, $\mathrm{Var}[\bar{\eta}^\dagger/\eta_C]
\simeq (t [I_{\rm er}''(\bar{\eta}^\dagger/\eta_C)])^{-1}$, and provides a measure of the precision of the machine operation at optimal performance, which should be read along the vertical axis.

As can be appreciated for the three curves in Fig.~\ref{fig:eraser_performance}, starting from an unbiased tape ($\delta_{\rm tape}=0$) and increasing the tape bias, we observe how both the erasure rate and the efficiency are enhanced until the maximum erasure rate is reached. Beyond this point, further increasing $\delta_{\rm tape}$ only raises the efficiency at the expense of a progressively smaller erasure rate. Eventually, the equilibrium point is approached, $\delta_{\rm tape}\rightarrow\delta_{\rm th}$, where the erasure rate vanishes and $\bar{\eta}^\dagger/\eta_C\rightarrow1$. Interestingly, close to the maximum erasure point the efficiency fluctuations are moderately reduced, while when approaching the equilibrium point, the efficiency fluctuations become increasingly large, specially as the temperature bias $\beta_{\rm h}/\beta_{\rm c}$ approaches $1$.

Indeed the thermal gradient controls the scale of the power-efficiency trade-off. Smaller values of $\beta_{\rm h}/\beta_{\rm c}$, corresponding to stronger thermal gradients, allows a substantially larger maximum in the erasure rates, but lower efficiencies (accompanied by also lower efficiency fluctuations). As $\beta_{\rm h}/\beta_{\rm c}$ increases, the maximum attainable erasure rate becomes smaller, while the maximum-power point moves up towards higher efficiencies (with also higher fluctuations). In the linear-response limit of the fridge, when $\beta_{\rm h}/\beta_{\rm c} \to 1$, the efficiency at maximum erasure rate approaches the universal value $\bar{\eta}^\dagger/\eta_C \rightarrow 1/2$ as for conventional thermal machines~\cite{vdB2009,VanDenBroeck2005}, indicated by the dotted line in Fig.~\ref{fig:eraser_performance}. In this limit the variance of the efficiency diverges.

\subsection{Optimal Refrigeration}
\begin{figure}[tbh]
    \centering
    \includegraphics[width=\columnwidth]{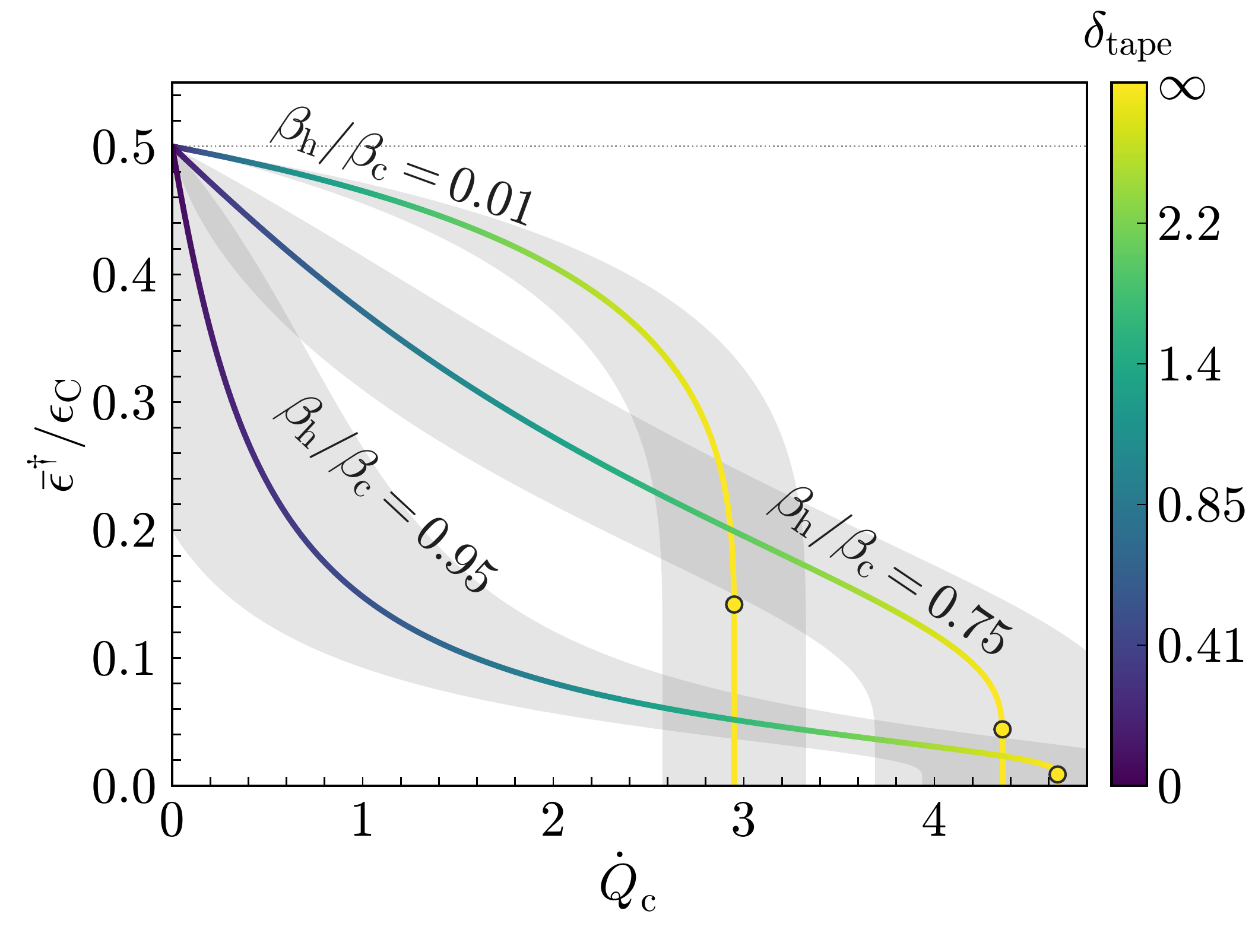}
    \caption{
    Optimal performance of the information-assisted refrigerator.
    For each tape bias $\delta_{\rm tape}$, the energy gap $\Delta E$
    is chosen to maximize the cooling power $\dot Q_{\rm c}$ at fixed
    bath temperatures, and the reduced COP
    $\bar{\epsilon}^\dagger/\bar{\epsilon}_{\rm C}$ is evaluated at the
    same operating point. Colors indicate $\delta_{\rm tape}$, while
    the three curves correspond to
    $\beta_{\rm h}/\beta_{\rm c}=0.01,\,0.75,$ and $0.95$.
    Gray bands represent fluctuations in the cold heat current along
    the horizontal axis. The dotted line marks the limiting value
    $\bar{\epsilon}^\dagger/\bar{\epsilon}_{\rm C}=1/2$ for weak driving.
    Yellow circles show the limiting reduced COP without the reset
    contribution as $\delta_{\rm tape}\to\infty$
    ($p_0^{\rm tape}\to1$), evaluated at the same optimal energy gaps.
    }
\label{fig:refrigerator_optimization_main}
\end{figure}

To characterize optimal refrigeration, we again fix the tape bias $\delta_{\rm tape}$ and maximize the heat extracted from the cold bath, $\dot Q_{\rm c}>0$ across the energy-temperature scale $\beta_{\rm c} \Delta E$. Optimal performance curves are then obtained when $\delta_{\rm tape}$ is varied, which are shown in Fig.~\ref{fig:refrigerator_optimization_main}. In contrast to the eraser optimization, here the gray shaded bands correspond to the fluctuations on the optimized heat current from the cold bath, as measured by ${\rm Var}[\dot{Q}_{\rm c}]$, and should be read in the horizontal axis~\footnote{Since in the refrigerator regime the fluctuations in the stochastic COP are trivially zero, we did not included them and instead used the current fluctuations to obtain a measure of the precision in the cooling operation}.

The curves in Fig.~\ref{fig:refrigerator_optimization_main} are obtained by maximizing the cooling power with respect to the energy gap $\Delta E$, at fixed tape bias and bath temperatures, and evaluating the COP at the same optimal operating point. Their behavior differs qualitatively from that of optimal erasure. Increasing $\Delta E$ increases the heat extracted per net transition, but also strengthens the opposing thermal bias and suppresses the net refrigeration current. The optimal gap balances these effects, producing a finite maximum cooling power for each value of $\delta_{\rm tape}$.

As the tape bias increases, the optimal gap approaches a finite value and the maximum cooling power saturates, while the reduced COP decreases towards zero. Consequently, the curves develop an almost vertical branch: increasingly biased input tapes produce little additional cooling power but incur an increasingly large informational cost. This behavior reflects the cost of restoring an almost pure incoming tape after its interaction with the machine~\footnote{The decrease of the COP is induced by the reset contribution in Eq.~\eqref{eq:Stape}. If only the tape entropy-change rate $\dot S_{\rm tape}$ were included as the resource, the corresponding COP would instead approach a finite constant as $\delta_{\rm tape}\rightarrow\infty$. The yellow circles indicate these limiting no-reset values, evaluated at the same optimal energy gaps.}. The asymptotic variance of the cold heat current also approaches a finite value, with a finite ratio to the mean current, consistently with the behavior of the signal-to-noise ratio in Fig.~\ref{fig:current_fluctuations}.

For larger temperature gradients, the reduced COP remains closer to its weak-driving value over a broader range of cooling powers. As the temperature gradient decreases, this region narrows and the COP falls more rapidly along the optimized branch. Unlike optimal erasure, however, a larger temperature gradient does not increase the attainable cooling power. Instead, the maximum cooling power increases as the gradient is reduced, while the Carnot-normalized COP vanishes in the limit $\beta_{\rm h}/\beta_{\rm c}\rightarrow1$.

At the low-power end, all curves approach $\bar{\epsilon}^{\dagger}/\epsilon_{\rm C}=1/2$, characteristic of the weak-driving, tight-coupling limit~\cite{Barato_2014,vdB2009,VanDenBroeck2005}, although their curvatures depend on $\beta_{\rm h}/\beta_{\rm c}$. For a large temperature gradient, the COP remains close to this limiting value even at appreciable cooling powers. For a small gradient, approaching the same limit requires a much stronger reduction of the cooling power. Along these branches, the energy gap tends to zero as $\delta_{\rm tape}\rightarrow0$, and the asymptotic variance of the cold heat current consequently vanishes.

An important difference with respect to the eraser regime, is that the refrigerator does not approach the Carnot COP at maximum cooling power, even when the temperature ratio is varied. Reversible refrigeration occurs only at the boundary where the the tape bias exactly compensates the thermal bias and the net current vanishes. However, this point cannot be reached when maximizing the cooling power with respect to $\Delta E$, which instead leads to a balance between the energy transported per transition against the remaining net current.

\section{Effects of quantum coherence}
\label{sec:coherence}

\begin{figure*}[t]
    \centering
    \includegraphics[width=\textwidth]{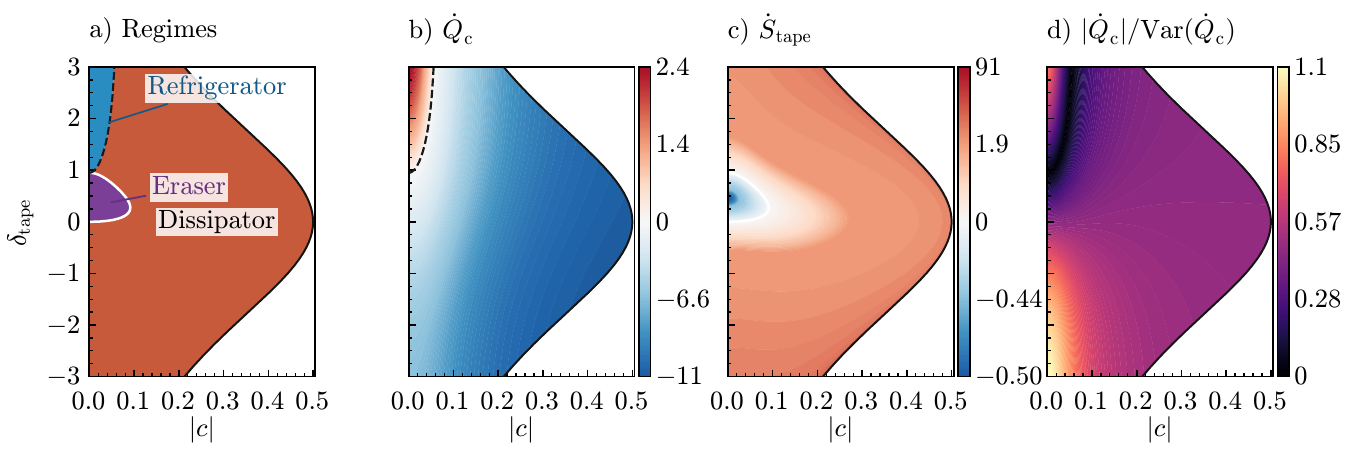}
    \caption{
    Effect of tape coherence on the stationary operation of the machine over
    the physically allowed $(|c|,p^{\rm tape}_0)$ domain.
    \textbf{(a)} Operating regimes.
    \textbf{(b)} Stationary cold heat current $\dot Q_{\rm c}$.
    \textbf{(c)} Complete informational contribution
    $\delta_{\rm tape}^c J_{\rm tape}^c$.
    \textbf{(d)} Cold-current precision
    $|\dot Q_{\rm c}|/\mathrm{Var}(\dot Q_{\rm c})$.
    The classical limit corresponds to $|c|=0$, while the outer boundary
    represents pure incoming tape states.
    }
    \label{fig:coherence_stationary}
\end{figure*}

Quantum coherence has attracted considerable attention in the study of
quantum thermal machines~\cite{Mitchison2019,latune2021roles}. It may modify 
energy transport~\cite{Mitchison2015}, improve thermodynamic performance~\cite{correa2014quantum,hardal2015superradiant,Hammam2021}, allow otherwise forbidden hybrid regimes of operation~\cite{Hammam2024}, or improve the precision of the currents~\cite{almanza2026}. In particular, autonomous
machines offer a natural setting in which coherence can be continuously
supplied by auxiliary systems as a genuinely nonclassical source of driving and converted into stationary transport without external time-dependent control~\cite{Manzano2019,Hammam2021}.

The analysis developed so far has deliberately focused on the classical setup in which the elements of the tape are classical bits of information, 
and hence it is fully characterized by the probabilities of being populated with $0$'s and $1$'s. We now extend this description to a quantum tape (or quantum information reservoir) by upgrading the tape bits to qubits, that may carry non-vanishing coherence in their computational basis $\{ \ket{0}, \ket{1} \}$. This provides a direct way to assess how coherence modifies the conversion between information and thermal resources. 

Accordingly, incoming qubits in the tape are characterized by the density operator
\begin{equation}
    \rho_{\rm tape}=
    \begin{pmatrix}
        p^{\rm tape}_0 & c\\
        c^* & p^{\rm tape}_1
    \end{pmatrix},
    \label{eq:coherent_tape}
\end{equation}
where the classical model is recovered for $c=0$. Here above $|c|^2\leq p^{\rm tape}_0p^{\rm tape}_1$ ensures a positive semi-definite density operator. The tape Hamiltonian is $ H_{\rm tape} = \Delta E \ket{1}\bra{1}$. Since $\delta_{\rm tape}=\ln(p^{\rm tape}_0/p^{\rm tape}_1)$, fixing the tape
populations is equivalent to fixing the bias $\delta_{\rm tape}$. The classical and quantum regimes can therefore be compared at the same population bias, isolating the contribution arising specifically from coherence (see Fig.~\ref{fig:coherence_stationary}).

The machine Hamiltonian can then be written as
    $H_{\rm m}=\Delta E\left(
    \ket{01}\!\bra{01}_{\rm m}
    +\ket{10}\!\bra{10}_{\rm m}\right)
    +2 \Delta E \ket{11}\!\bra{11}_{\rm m}$,
with degenerate levels $\ket{01}_{\rm m}$ and $\ket{10}_{\rm m}$, and the repeated interactions (or collisions) with the qubits of the tape are governed by an energy-preserving interaction Hamiltonian:
\begin{equation}
\label{eq:H_int}
H_{\rm int} = g \left(
    \ket{01}\!\bra{10}_{\rm m} \ket{0}\bra{1}
    +\ket{10}\!\bra{01}_{\rm m}  \ket{1}\bra{0}\right),
\end{equation}
where $g = \phi/\tau$ is the coupling strength and $\tau$ denotes the interaction time. The above interaction Hamiltonian allows the coherent exchange between the degenerate levels of the machine and the tape qubits. 

In the weak-collision limit, a quantum master equation can be derived that describes the evolution of the machine subjected to repeated interactions with the tape qubits (see App.~\ref{app:coherent_dynamics}):
\begin{equation}
    \dot\rho_{\rm m}=-i[H_{\rm m}+V,\rho_{\rm m}]
    +\sum_{\alpha=h,c,\mathrm{tape}}\mathcal L_\alpha(\rho_{\rm m}),
    \label{eq:coherent_master_main}
\end{equation}
where $V= r\phi \left(c\,\ket{10}\!\bra{01}_{\rm m} +c^*\,\ket{01}\!\bra{10}_{\rm m}\right)$ is a driving-like term induced by the presence of coherence in the tape qubits~\cite{Manzano2019,rodrigues2019thermodynamics}, and $\mathcal L_i$ for $i={\rm h, c, tape}$ describe incoherent transitions leading to the same jump-like dynamics as in the classical model. The corresponding heat currents in the NESS can be defined as 
\begin{equation}
    \dot Q_\alpha=
    \operatorname{Tr}\left[
    H_{\rm m}\,\mathcal L_\alpha(\rho_{\rm m}^{\rm ss})
    \right],
    \qquad \alpha=h,c,
    \label{eq:coherent_heat_current}
\end{equation}
where $\rho_{\rm m}^{\rm ss}$ denotes the machine density operator in the long-time run. 

Similarly, we are interested in the changes of the von Neumann entropy of the tape qubits in the NESS, related to the change in the qubit state during the collisions.
In order to obtain it. We use the spectral decomposition of $\rho_{\rm tape} =\lambda_+\Pi_+ + \lambda_-\Pi_-$, where $\lambda_\pm$ and $\Pi_\pm$ denote, respectively, its eigenvalues and corresponding eigenprojectors. In particular, the eigenvalues read $\lambda_\pm =(1\pm R_q)/2$, with $R_q= \sqrt{(2p_0^{\rm tape}-1)^2 +4|c|^2}$. The corresponding quantum tape current can be then written as
\begin{equation}
    J_{\rm tape}^{\rm q}
    \equiv
    \frac{r}{2}
    \operatorname{Tr}
    \left[
        ( \rho_{\rm tape}' - \rho_{\rm tape})
        (\Pi_+-\Pi_-)
    \right],
    \label{eq:coherent_bias_current}
\end{equation}
where $\rho_{\rm tape}^\prime$ is the state of the qubits after interaction with the machine, as detailed in appendix~\ref{app:coherent_dynamics}.

The rate of change in the von Neumann entropy of the tape can then be written, similarly to the classical case, as 
\begin{equation}
    \dot S_{\rm tape}
    =
    -\delta_{\rm tape}^{\rm q} J_{\rm tape}^{\rm q} - r D(\rho_{\rm tape}'\Vert\rho_{\rm tape}),
    \label{eq:quantum_information_cost}
\end{equation}
with $\delta_{\rm tape}^{\rm q} \equiv  \ln(\lambda_+/\lambda_-)$ a quantum version of the tape bias in Eq.\eqref{tape_bias}. For zero coherence ($c=0$), we have $J_{\rm tape}^{\rm q} = J_{\rm tape}$ and $\delta_{\rm tape}^{\rm q}=\delta_{\rm tape}$, so that Eq.~\eqref{eq:Stape} is recovered.

We are interested in the impact of coherence on the available regimes of operation, the magnitude of the stationary currents, and the fluctuations in the machine, as illustrated in Fig.~\ref{fig:coherence_stationary}(a)-(d). Importantly, our results show that only the modulus of $c$ has an impact in the thermodynamic quantities of interest, while its phase is irrelevant. In Fig.~\ref{fig:coherence_stationary} it can be appreciated that, despite all the three regimes of operation of the machine can be found for non-zero coherence, increasing the absolute value of $|c|$ results predominantly in an enlargement of the
dissipative region rather than in the emergence of a new coherence-assisted useful operating mode.

These detrimental effects can be understood from the fact that the role of tape coherence on the thermal machine dynamics is qualitatively different from the one played by tape populations. While the later is able to bias transitions between the degenerate states $\ket{01}_{\rm m}$ and $\ket{10}_{\rm m}$, the former induces a coherent exchange without any preferred direction. Since the thermal gradient of the reservoirs naturally favors the spontaneous flow of heat from hot to cold reservoirs, increasing $|c|$ at fixed $\delta_{\rm tape}$ progressively weakens the ability of the population bias to oppose this tendency. As a consequence, the dissipative regime expands at the expense of both useful operating modes [Fig.~\ref{fig:coherence_stationary}(a)], while the cold heat current decreases until it takes negative values [Fig.~\ref{fig:coherence_stationary}(b)] and refrigeration is lost. 
We also notice that, although increasing the coherence $|c|$ leads to greater values of the quantum bias $\delta_{\rm tape}^{\rm q}$, the additional asymmetry does not translate into useful thermodynamic action: even if the tape becomes increasingly pure in its eigenbasis, the induced coherent dynamics predominantly enhances mixing within the machine degenerate subspace. This in turn leads to higher values of $\dot{S}_{\rm tape}$ which tends to spoil the eraser regime. This effect becomes especially pronounced close to the pure-state boundary, where $\lambda_-\rightarrow0$ and $\dot{S}_{\rm tape}$ becomes very large [Fig.~\ref{fig:coherence_stationary}(c)].

Finally, coherence also modifies the precision of stationary transport, beyond its effect on the average currents. States with the same populations, and hence the same classical bias $\delta_{\rm tape}$, can display substantially different fluctuations when $|c|$ is varied. The corresponding changes in the signal-to-noise ratio,
$|\dot Q_{\rm c}|/\mathrm{Var}(\dot Q_{\rm c})$, are shown in Fig.~\ref{fig:coherence_stationary}(d). 
Over most of the accessible domain, the signal-to-noise ratio remains relatively small and never increases with respect to the corresponding values for $|c|=0$, indicating that the changes in the stationary heat currents are accompanied by sizeable fluctuations that worsen the precision.

\section{Conclusions and discussion}
\label{sec:conclusion}

We have presented an autonomous information machine model acting on a tape of energetically degenerate two-level systems (bits), which enables conversion between thermal and informational resources. The main thermodynamic currents in the NESS are controlled by the competition between a thermal bias induced by the baths, and the tape bias (frequency of $0$'s with respect to $1$'s in the incoming tape bits), which determines whether the machine operates as an information eraser driven by heat, an information-powered refrigerator, or a dissipator.

For both the eraser and the refrigerator regimes, we established suitable efficiencies based on two different second-law inequalities [Eqs.~\eqref{eq:second_law_eraser} and \eqref{eq:second_law_fridge}]. These correspond to two different definitions of entropy production, that arise depending on the assumed level of control~\cite{Manzano2018,ptaszynski2019entropy}. Here these are related to the different operational roles of the tape in the two regimes: erasure retains the outgoing tape as an useful product, leading to efficiency in Eq.~\eqref{eq:eraser_efficiency} which is bounded by Carnot's efficiency. On the other hand, refrigeration uses a low-entropy tape as a resource and hence the cost of restoring the initial state of the tape bits becomes relevant, leading to COP in Eq.~\eqref{eq:refrigerator_cop}, bounded by power-driven Carnot's COP. 

Using full counting statistics and large deviation theory, we characterized current and efficiency fluctuations in the long-time limit. Transport precision is constrained by the mismatch between thermal and informational biases through the TUR [Eq.~\eqref{eq:tur_ratio_heat}], which becomes tight close to the equilibrium point $\delta_{\rm tape} = \delta_{\rm th}$, when approaching the linear response regime~\cite{almanza2026}. However, we found that the currents variance remain finite when their average values vanish at equilibrium, so approaching reversible operation entails increasingly poor precision, while it is enhanced far from equilibrium. In the weak-coupling regime between the machine and the tape, we also identified a characteristic point of operation inside the refrigerator regime at which the leading heat-current variance becomes independent of the temperature gradient.

Efficiency fluctuations reveal qualitative differences between the two useful operating regimes. For erasure, the reversible efficiency need not be the least probable value, departing from the standard result for conventional thermal machines~\cite{Verley2014a}. For refrigeration, tight coupling between the heat currents and the informational input suppresses COP fluctuations in the long-time limit. Therefore, our results show that while the machine would be able to function at always at its macroscopic efficiency in the refrigerator regime, the same is not true for erasing information. This contrast reflects again the different nature of the relevant informational quantities entering the efficiencies in the two regimes.

The above results motivate the optimization of the machine for working at maximum (erasing/cooling) power. 
As expected, strong thermal gradients increase the maximum erasure rate, but reduce the associated efficiency at maximum power, which remains bounded by $\bar{\eta} = \eta_{\rm C}/2$ when approaching the linear response regime, in agreement with Refs.~\cite{Barato_2014,vdB2009,VanDenBroeck2005}. Efficiency fluctuations become moderately low at maximum power, specially for high temperature gradients. For refrigeration, instead, maximum cooling power is favored for small thermal grandients and high tape biases,  while the reset cost of an increasingly ordered tape drives the COP towards zero. In the tight-coupling limit, the COP at maximum cooling power again recovers $\bar{\epsilon} = \epsilon_{\rm C}/2$.

Extending the model to allow quantum coherence in the tape we showed how it modifies transport through the introduction an extra coherent exchange between the machine’s degenerated levels. At fixed tape populations, increasing coherence predominantly favors spontaneous heat transport, expands the dissipative region, and narrows the ranges of useful erasure and refrigeration in the parameter regime studied. It also generally reduces heat-current precision and decreases efficiency. These findings thus demonstrate that the presence of coherence does not necessarily improve performance, contrary to other models using energetic coherence to improve the performance of quantum thermal machines~\cite{correa2014quantum,hardal2015superradiant,Hammam2021}.

The derivation of the present model based on a repeated-interaction scheme suggests a route to experimental tests using either a micro-maser setup~\cite{meschede1985one,rempe1987observation} or programmable qubit platforms, including trapped-ion
systems~\cite{Schindler2013} and superconducting
processors~\cite{Cattaneo2023}. In the first case, flying atoms can play the role of the informational tape~\cite{gleyzes2007quantum,najera2020autonomous}, while the machine could be engineered trapping atoms inside the microwave cavity subjected to engineered dissipation using lasers. In the second case, sequentially prepared auxiliary systems could emulate the informational tape, while engineered dissipative channels reproduce the thermal reservoirs. Such implementations would allow the predicted operating regimes and fluctuations to be tested.

Natural extensions of the present setup include information reservoirs bearing correlations between the input symbols and thermal machines with multiple thermodynamic cycles. Correlations between successive tape elements may modify the dynamics and thermodynamic currents at both the average and fluctuation levels, while may allow performing more complex computational tasks over the tape~\cite{Boyd2018,Loomis2020,Jurgens2020}. However recovering the full computational power of Turing machines might require adding and extra tape~\cite{wolpert2019stochastic}. In machines with multiple cycles, informational reservoirs could also provide additional means of controlling competing transport processes and enabling further operating regimes. The detrimental effects of coherence observed here also motivate a broader study of alternative machine architectures and coupling mechanisms. Such investigations may identify conditions under which coherence can support directed transport and improve useful thermodynamic performance.

% ============================================================

\acknowledgments{We thank Rosa L\'opez for fruitful discussions. We acknowledge support from the CoQuSy project PID2022-140506NB-C21 and C22 and the Mar\'ia de Maeztu project CEX2021-001164-M for Units of Excellence, funded by MICIU/AEI/10.13039/501100011033/FEDER, UE. LT acknowledges CSIC JAE-PRE program [grant JAEPRE23-26].}

\appendix

\section{Master equation details}
\label{app:classical_generator}
In this appendix, we provide the explicit form of the rate matrix $\mathbb W$ governing the Markovian dynamics of the machine, and provide details on its derivation. It contains contributions from the two thermal reservoirs locally acting on each two-level subsystem of the machine and an extra contribution from the interactions with the tape. 
The role of the thermal reservoirs is just to induce local transitions in each two-level subsystem with poissoninan statistics. Each allowed excitation or relaxation ($\uparrow \downarrow$) occurs with a probability during an
infinitesimal interval $[t, t+dt]$ reading $\Gamma_i^{\uparrow \downarrow}dt + O(dt^2)$ for $i={\rm h, c}$, where $\Gamma_i^{\uparrow \downarrow}$ are the transition rates verifying local detailed balance at the respective bath temperature [Eq.~\eqref{eq:rates}]. 

The tape contribution can be obtained by assuming the arrival of fresh, independent bits of the tape at random times. Similarly to the case of a thermal bath, we consider an interaction between the machine and a bit of the tape to occur with a probability $r\,dt + O(dt^2)$ during the interval $[t, t+ dt]$. Once an interaction is verified, it allows the joint exchange $(\ket{01}_{\rm m},\ket{0})\rightleftarrows(\ket{10}_{\rm m},\ket{1})$, where $\ket{0}$ and $\ket{1}$ denote the states of the incoming bit, with (conditional) probability $\phi^2$ given by the effective interaction strength $\phi \ll 1$. Therefore the probability of a marginal transition $\ket{01}_{\rm m} \rightarrow \ket{10}_{\rm m}$ in the machine reads $\Gamma_{\rm tape}^{\uparrow} = r \phi^2 p_0$, which is proportional to the probability of the tape to be in state $\ket{0}$. Analogously the probability of a transition $\ket{10}_{\rm m} \rightarrow \ket{01}_{\rm m}$, denoted $\Gamma_{\rm tape}^{\downarrow} = r \phi^2 p_1$, is proportional to $p_1$.

In the column vector convention $\dot{\mathbf p}_{\rm m}=\mathbb W \,\mathbf p_{\rm m}$, with $\bm{p}_{\rm m} = (p_{00},p_{01},p_{10},p_{11})^T$, the generator reads
\begin{widetext}
\begin{equation}
\mathbb W=
\left(
\begin{array}{cccc}
-(\Gamma_{\rm c}^\uparrow+\Gamma_{\rm h}^\uparrow)
&\Gamma_{\rm c}^\downarrow&\Gamma_{\rm h}^\downarrow&0\\[2mm]
\Gamma_{\rm c}^\uparrow
&-(\Gamma_{\rm c}^\downarrow+\Gamma_{\rm h}^\uparrow+\Gamma_{\rm tape}^\uparrow)
&\Gamma_{\rm tape}^\downarrow&\Gamma_{\rm h}^\downarrow\\[2mm]
\Gamma_{\rm h}^\uparrow&\Gamma_{\rm tape}^\uparrow
&-(\Gamma_{\rm h}^\downarrow+\Gamma_{\rm c}^\uparrow+\Gamma_{\rm tape}^\downarrow)
&\Gamma_{\rm c}^\downarrow\\[2mm]
0&\Gamma_{\rm h}^\uparrow&\Gamma_{\rm c}^\uparrow
&-(\Gamma_{\rm h}^\downarrow+\Gamma_{\rm c}^\downarrow)
\end{array}
\right),
\label{eq:classical_generator}
\end{equation}
\end{widetext}
Here and in the following, we order the machine states as $\{\ket{00}_{\rm m},\ket{01}_{\rm m},\ket{10}_{\rm m},\ket{11}_{\rm m}\}$, with the first and second entries in the bra-kets referring to the two-level systems coupled to hot and cold reservoirs, respectively. In the column-vector convention, each off-diagonal element of $\mathbb W$ is the transition rate from the state labeling its column to that labeling its row. Each diagonal element is minus the total escape rate from that state, ensuring that every column sums to zero and probability is conserved.

In the long time run the dynamics become stationary and the populations satisfy $\mathbb W \, \mathbf p_{\rm m}^{\rm ss}=0$, which can be solved.
They read:
\begin{equation}
\begin{aligned}
p_{00}^{\rm ss}
&=\frac{
(\xi^\uparrow+\xi^\downarrow)
\Gamma_{\rm h}^\downarrow\Gamma_{\rm c}^\downarrow
+\xi^\downarrow\mathbb \Gamma(\Gamma_{\rm h}^\downarrow,\Gamma_{\rm c}^\downarrow)
}{\mathcal Z},
\\[2mm]
p_{01}^{\rm ss}
&=\frac{
(\xi^\uparrow+\xi^\downarrow)
\Gamma_{\rm h}^\downarrow\Gamma_{\rm c}^\uparrow
+\Gamma_{\rm tape}^\downarrow\xi^\uparrow\xi^\downarrow
}{\mathcal Z},
\\[2mm]
p_{10}^{\rm ss}
&=\frac{
(\xi^\uparrow+\xi^\downarrow)
\Gamma_{\rm h}^\uparrow\Gamma_{\rm c}^\downarrow
+\Gamma_{\rm tape}^\uparrow\xi^\uparrow\xi^\downarrow
}{\mathcal Z},
\\[2mm]
p_{11}^{\rm ss}
&=\frac{
(\xi^\uparrow+\xi^\downarrow)
\Gamma_{\rm h}^\uparrow\Gamma_{\rm c}^\uparrow
+\xi^\uparrow\mathbb \Gamma(\Gamma_{\rm c}^\uparrow,\Gamma_{\rm h}^\uparrow)
}{\mathcal Z},
\end{aligned}
\label{eq:classical_stationary_populations}
\end{equation}
where we introduced the auxiliary rates,
$\xi^\uparrow=\Gamma_{\rm h}^\uparrow+\Gamma_{\rm c}^\uparrow$,
$\xi^\downarrow=\Gamma_{\rm h}^\downarrow+\Gamma_{\rm c}^\downarrow$,

the function $\mathbb \Gamma(x,y)=\Gamma_{\rm tape}^\uparrow x
              +\Gamma_{\rm tape}^\downarrow y,
$
and the normalization factor: 
\begin{equation}
\begin{aligned}
\mathcal Z={}&(\xi^\uparrow+\xi^\downarrow)
\Big[
(\Gamma_{\rm h}^\uparrow+\Gamma_{\rm h}^\downarrow)
(\Gamma_{\rm c}^\uparrow+\Gamma_{\rm c}^\downarrow)
+\mathbb \Gamma(\Gamma_{\rm h}^\downarrow,\Gamma_{\rm c}^\downarrow)
\Big]
\\
&+\xi^\uparrow\mathbb \Gamma
\big(\Gamma_{\rm c}^\uparrow+\Gamma_{\rm c}^\downarrow,\,
     \Gamma_{\rm h}^\uparrow+\Gamma_{\rm h}^\downarrow\big).
\end{aligned}
\label{eq:classical_normalization}
\end{equation}
The stationary tape current then becomes, using Eqs.~\eqref{eq:classical_stationary_populations}:
\begin{equation}
J_{\rm tape}
=\frac{(\xi^\uparrow+\xi^\downarrow)
\left(
\Gamma_{\rm tape}^\downarrow\Gamma_{\rm h}^\uparrow\Gamma_{\rm c}^\downarrow
-\Gamma_{\rm tape}^\uparrow\Gamma_{\rm h}^\downarrow\Gamma_{\rm c}^\uparrow
\right)}{\mathcal Z},
\label{eq:classical_stationary_current}
\end{equation}
and the currents can be directly shown to satisfy $J_h=-J_c=J_{\rm tape}$, and $\dot Q_h=-\dot Q_c=\Delta E\,J_{\rm tape}$.

\section{Tape dynamics and entropy changes}
\label{app:tape_entropy}
 
The informational machine is designed so that each transition between the machine states $\ket{10}_{\rm m} \to \ket{01}_{\rm m}$ is responsible of a bit flip in the tape from $1$ to $0$, and, similarly, a transition $\ket{01}_{\rm m} \to \ket{10}_{\rm m}$ triggers a flip in the tape bit from $0$ to $1$. Since fresh bits in the tape interact with the machine at rate $r$, once the machine has reached the steady state, the probability density of the tape bits after interaction change as
\begin{equation}
p_0^{\prime\,\rm tape}=p_0^{\rm tape}+\frac{J_{\rm tape}}{r},
\label{eq:tape_outgoing_probabilities}
\end{equation}
where $J_{\rm tape}=\Gamma_{\rm tape}^\downarrow p_{10}^{\rm ss}-\Gamma_{\rm tape}^\uparrow p_{01}^{\rm ss}$ is the net probability for a transition $\ket{10}_{\rm m} \to \ket{01}_{\rm m}$ in the machine at the NESS, and $p_1^{\prime \, \rm tape} = 1- p_0^{\prime \, \rm tape}$.

The change in Shannon entropy $H({\bm p})=-\sum_i p_i\ln p_i$ of a tape bit due to the interaction with the machine then reads:
\begin{align} \label{eq:tape_entropy_identity}
&H({\bm p}_{\rm tape}')-H({\bm p}_{\rm tape}) = \\
&\, =-\sum_i
\left(p_i^{\prime\,\rm tape}-p_i^{\rm tape}\right)
\ln p_i^{\rm tape}
-D({\bm p}_{\rm tape}'\Vert {\bm p}_{\rm tape}), \nonumber
\end{align}
where, in the second line, we added and subtracted the term $-\sum_i p_i^{\prime\,\rm tape}\ln p_i^{\rm tape}$ to recover the Kullback-Leibler divergence $D({\bm p}_{\rm tape}'\Vert {\bm p}_{\rm tape}) \equiv \sum_i p_i^{\prime\,\rm tape} \ln(p_i^{\prime\,\rm tape}/p_i^{\rm tape})$ between the tape distributions before and after interaction.

Substituting Eq.~\eqref{eq:tape_outgoing_probabilities} into Eq.~\eqref{eq:tape_entropy_identity} yields the equality
\begin{equation}
-\delta_{\rm tape}J_{\rm tape}
=\dot S_{\rm tape}
+rD(p_{\rm tape}'\Vert p_{\rm tape}),
\label{eq:tape_entropy_balance}
\end{equation}
which is used to reach the second line in the second-law inequality \eqref{eq:second_law_fridge}.
The extra term $D({\bm p}_{\rm tape}'\Vert {\bm p}_{\rm tape})$ can be interpreted as the 
entropy production due to the reset of the outgoing tape bits to its incoming distribution ${\bm p}_{\rm tape}$ 
under free relaxation (homogeneization) dynamics~\cite{Manzano2018}.

At the stochastic level, in a single interaction where the state of the bit changes from state $x$ to $y$, we introduce the change in the surprisal of the tape bit as
\begin{equation}
\label{eq: increment}
\Delta s_{xy}
=\ln\frac{p_x^{\rm tape}}{p_y^{\prime\,\rm tape}},
\end{equation}
whose average over realizations recovers the Shannon entropy changes, $\langle\Delta s_{xy}\rangle
=H({\bm p}_{\rm tape}')-H({\bm p}_{\rm tape})$. Similarly we also consider the stochastic distinguishability associated to the tape bit state after interaction
\begin{equation}
    d_y
=\ln\frac{p_y^{\prime\,\rm tape}}{p_y^{\rm tape}},
\end{equation}
which recovers the reset term $\langle d_y\rangle
=D({\bm p}_{\rm tape}'\Vert{\bm p}_{\rm tape})$.

\section{Current statistics}
\label{app:current_statistics}

Here we provide the details of the full counting statistics (FCS)
formalism~\cite{Esposito2009b} that we used to determine current and
efficiency fluctuations, extending the standard framework to allow its application to an information reservoir in the form of a tape. 

We are interested in the stochastic dynamics of the integrated heat currents from the hot and cold baths, $q_{\rm h}(t)=\Delta E X_{\rm h}(t)$ and $q_{\rm c}(t)=\Delta E X_{\rm c}(t)$ respectively.
The accumulated changes in the Shannon entropy of the tape, $\Delta s_{\rm tape}(t) $, and the accumulated changes of the reset contribution, which we denote $\Delta d_{\rm tape}(t)$. For each of these four stochastic variables, we introduce a conjugate counting field:
\begin{equation}
\boldsymbol\lambda=(\lambda_{\rm h},\lambda_{\rm c},\lambda_{\rm s},\lambda_{\rm d}),
\label{eq:counting_fields}
\end{equation}
which are introduced into the continuous-time tilted generator:
\begin{equation}
\mathbb W_{\boldsymbol\lambda}
=\mathbb W_h(\lambda_h)+\mathbb W_c(\lambda_c) + \mathbb W_{\rm tape}(\lambda_{\rm s}, \lambda_{\rm d}).
\label{eq:classical_tilted_generator}
\end{equation}
Here above $\mathbb{W}_{\rm h}(\lambda_{\rm h})$ and $\mathbb{W}_{\rm c}(\lambda_{\rm c})$ are the generators of the local transitions in each subsystem driven by the hot and cold baths respectively. They can be written as:
\begin{equation}
\mathbb W_{\rm h}(\lambda_{\rm h})
=L_{\rm h}(\lambda_{\rm h})\otimes \mathbb{1}_2,
\qquad
\mathbb W_{\rm c}(\lambda_{\rm c})
=\mathbb{1}_2\otimes L_{\rm c}(\lambda_{\rm c}),
\end{equation}
where $\mathbb{1}_2$ is the two-dimensional identity matrix and
\begin{equation}
L_i(\lambda_i)=
\begin{pmatrix}
-\Gamma_i^\uparrow
&\Gamma_i^\downarrow e^{-\lambda_i\Delta E}\\
\Gamma_i^\uparrow e^{\lambda_i\Delta E}
&-\Gamma_i^\downarrow
\end{pmatrix},
\qquad i\in\{h,c\}.
\end{equation}
In the tilted generator, the transitions in the local subsystems were weighted by factors proportional to the counting fields $e^{\pm\lambda_i\Delta E}$, used to count the transitions; while the escape rates remained unchanged. On the other hand, the contribution from the tape dynamics can be written as
$W_{\rm tape}(\lambda_{\rm s}, \lambda_{\rm d}) = r\bigl(R_{\lambda_s,\lambda_d}- \mathbb{1}_4\bigr)$ in terms of the following tilted map associated to a single tape-machine interaction:
\begin{widetext}
\begin{equation}
R_{\lambda_s,\lambda_d}=
\begin{pmatrix}
p_0F_{00}+p_1F_{11} &0&0&0\\[1mm]
0&p_0(1-\phi^2)F_{00}+p_1F_{11}&p_1\phi^2F_{10}&0\\[1mm]
0&p_0\phi^2F_{01}&p_0F_{00}+p_1(1-\phi^2)F_{11}&0\\[1mm]
0&0&0&p_0F_{00}+p_1F_{11}
\end{pmatrix},
\label{eq:classical_tilted_collision_map}
\end{equation}
\end{widetext}
with factors $F_{xy}(\lambda_{\rm s}, \lambda_{\rm d}
)=\exp(\lambda_s \Delta s_{xy}+\lambda_d d_y)$ proportional to the counting fields $\lambda_{\rm s}$ and $\lambda_{\rm d}$. The scope of the two different counting fields ($\lambda_{\rm s}$ and $\lambda_{\rm d}$) is to track separate records of the entropy change contributions (surprisal) and reset terms (distinguishability). At zero counting fields, we have $\mathbb W_{\rm tape}=r(R_{0,0}-\mathbb{1}_4)$ recovering the corresponding contribution to the rate matrix of the tape transition rates  $\Gamma_{\rm tape}^\uparrow=r\phi^2p_0^{\rm tape}$ and $\Gamma_{\rm tape}^\downarrow=r\phi^2p_1^{\rm tape}$.

The scaled cumulant generating function (SCGF) is the dominant
eigenvalue of the tilted generator $\mathbb W_{\boldsymbol\lambda}$ and can be written as
\begin{equation}
\Psi(\boldsymbol\lambda)
=\lim_{t\to\infty}\frac{1}{t}
\ln\left\langle
\exp\bigl[\boldsymbol\lambda\cdot\boldsymbol Y(t)\bigr]
\right\rangle,
\label{eq:classical_joint_scgf}
\end{equation}
where we introduced the vector of stochastic variables:
\begin{equation}
\boldsymbol Y(t)=
\begin{pmatrix}
\Delta E X_h(t)\\
\Delta E X_c(t)\\
\Delta s_{\rm tape}(t)\\
\Delta d_{\rm tape}(t)
\end{pmatrix}.
\end{equation}
Its derivatives at zero fields generate the stationary mean
rates and asymptotic covariance rates. In particular,
\begin{equation}
\left.\partial_{\lambda_i}\Psi\right|_{\boldsymbol\lambda=0}
=\lim_{t\to\infty}\frac{\langle Y_i(t)\rangle}{t},
\label{eq:classical_scgf_first_derivative}
\end{equation}
so that for $i={\rm h, c}$ we obtain

$\dot Q_h$ and $\dot Q_c$ directly.

The first derivatives of the SCGF with respect to $\lambda_{\rm s}$ and $\lambda_{\rm d}$ also recover the mean fluxes for the main informational contributions:
\begin{align}
\left.\partial_{\lambda_s}\Psi\right|_{\mathbf 0}
&=r\sum_{x,y}P^{\rm ss}_{xy}
\ln\frac{p_x^{\rm tape}}{p_y^{\prime\,\rm tape}}
=\dot S_{\rm tape},
\\\left.\partial_{\lambda_d}\Psi\right|_{\mathbf 0}
&=r\sum_{x,y}P^{\rm ss}_{xy}
\ln\frac{p_y^{\prime\,\rm tape}}{p_y^{\rm tape}}
=rD\bigl(p^{\prime\,\rm tape}\Vert p^{\rm tape}\bigr),
\end{align}
where we denoted $P^{\rm ss}_{xy}$ the joint probability of an incoming tape bit in state $x$ and an outgoing bit in state $y$ after interaction (in the NESS), obtained from the kernel of $R_{\lambda_{\rm s}, \lambda_{\rm d}}$ in Eq.~\eqref{eq:classical_tilted_generator} when the counting fields are set to zero. To reach the last equalities above we used the marginalization,
$\sum_yP^{\rm ss}_{xy}=p_x^{\rm tape}$ and $\sum_xP^{\rm ss}_{xy}=p_y^{\prime\,\rm tape}$. This result confirms that the counting method employed here reproduces correctly both the Shannon entropy changes in the tape and the reset contribution in Eq.~\eqref{eq:Stape}, separately.

However, for assessing the tape probability current statistics it is convenient to count instead $X_{\rm tape}(t)$ using a single field, $\lambda_{\rm tape}$, with the transition $\ket{10}\to\ket{01}$ taken as generating a positive count. The corresponding tape-machine tilted interaction map reads
\begin{equation}
R_\chi=
\begin{pmatrix}
1&0&0&0\\
0&1-p_0\phi^2&p_1\phi^2e^{\lambda_{\rm tape}}&0\\
0&p_0\phi^2e^{-\lambda_{\rm tape}}&1-p_1\phi^2&0\\
0&0&0&1
\end{pmatrix}.
\label{eq:classical_current_tilted_map}
\end{equation}
In this case, writing $\mathbb W_{\lambda_{\rm tape}}=\mathbb W_h+\mathbb W_c+r(R_{\lambda_{\rm tape}}-I)$,
we obtain that the average current and its variance are given by
\begin{equation}
J_{\rm tape}=\Psi_{\rm tape}'(0),
\qquad
\mathrm{Var}(J_{\rm tape})=\Psi_{\rm tape}''(0),
\label{eq:classical_current_cumulants}
\end{equation}
where $\mathrm{Var}(J_{\rm tape})$ denotes the asymptotic variance
rate $\lim_{t\to\infty}\mathrm{Var}[X_{\rm tape}(t)]/t$.

The two versions of the counting formalism (either using the two fields $\lambda_{\rm s}$ and $\lambda_{\rm d}$ or the global field $\lambda_{\rm tape}$) are consistent. Indeed, setting
$\lambda_{\rm s}=\lambda_{\rm d}=\lambda_{\rm tape}$ gives
$s_{00}+d_0=s_{11}+d_1=0$ together with $s_{01}+d_1=\delta_{\rm tape}$ and $s_{10}+d_0=-\delta_{\rm tape}$. Consequently, we obtain the identity for the accumulated currents
\begin{equation}
\Delta s_{\rm tape}(t)+\Delta d_{\rm tape}(t)
=-\delta_{\rm tape}X_{\rm tape}(t).
\label{eq:classical_trajectory_tape_identity}
\end{equation}
which is the stochastic counterpart of Eq.\eqref{eq:second_law_fridge}.

Moreover, the conservation of excitations between the hot and cold reservoirs gives in the time asymptotic limit:
\begin{equation}
X_c\approx X_{\rm tape},
\qquad
X_h \approx X_{\rm tape}.
\label{eq:classical_trajectory_conservation}
\end{equation}
The differences between the
integrated currents involve only bounded boundary terms.
For the joint SCGF of the three probability currents,
this implies
\begin{equation}
\Psi_J(\lambda_{\rm c},\lambda_{\rm h},\lambda_{\rm tape})
=\Psi_{\rm tape}(\lambda_{\rm tape}+\lambda_{\rm h}-\lambda_{\rm c}).
\label{eq:classical_current_scgf_locking}
\end{equation}
Therefore the relation $J_h=-J_c=J_{\rm tape}$ derived in App.~\ref{app:classical_generator} extends to
the asymptotic current cumulants. In particular,
\begin{align}
\mathrm{Var}(J_c)
&=\mathrm{Var}(J_h)=\mathrm{Var}(J_{\rm tape}),\nonumber\\
\mathrm{Var}(\dot Q_i)
&=\Delta E^2\,\mathrm{Var}(J_{\rm tape}),
\qquad i\in\{h,c\}.
\label{eq:classical_variance_locking}
\end{align}
The same identity fixes all higher extensive cumulants and cross-cumulants, with the signs determined by the current orientations.

\section{Variance crossing point}
\label{app:fluctuation_crossing}

We now consider the regime of slow collisions (with respect to the dynamics induced by the thermal baths),
 $r\phi^2\ll \min(\Gamma^\uparrow_{\rm h}+\Gamma^\downarrow_{\rm h},\Gamma^\uparrow_{\rm c}+\Gamma^\downarrow_{\rm c})$, where a compact expression for the variance of the fluxes can be obtained analytically. We notice that this separation of time-scales is distinct from the weak-collision condition $\phi\ll1$, which is always verified. In the extreme case $r\phi^2=0$, the stationary machine distribution is the product thermal state, with $p_{01}^{\rm eq}=\left(1-\gamma^\uparrow_{\rm h} \right)\gamma^\uparrow_{c}$ and
$p_{10}^{\rm eq}=\gamma^\uparrow_{\rm h}(1-\gamma^\uparrow_{\rm c})$, where $\gamma^\uparrow_{\rm i}=\Gamma^\uparrow_{\rm i}/(\Gamma^\uparrow_{\rm i}+\Gamma^\downarrow_{\rm i})$.
First-order perturbation theory for the dominant eigenvalue
of the current-tilted generator gives
\begin{align}
\Psi_{\rm tape}(\lambda_{\rm tape})
={}&r\phi^2p_1^{\rm tape}p_{10}^{\rm eq}(e^{\lambda_{\rm tape}}-1) \\
&+r\phi^2p_0^{\rm tape}p_{01}^{\rm eq}(e^{-\lambda_{\rm tape}}-1)
+O\!\left(\frac{r^2\phi^4}{k_{\min}}\right). \nonumber
\label{eq:classical_weak_coupling_scgf}
\end{align}
The leading statistics are thus those of two independent
rare Poisson transfers. The resulting variance rate is
\begin{align}
\mathrm{Var}(J_{\rm tape})
={}&r\phi^2\bigl[
 p_0^{\rm tape}\gamma^\uparrow_{\rm c}
 +\gamma^\uparrow_{\rm h}(1-\gamma^\uparrow_{\rm c}-p_0^{\rm tape})\bigr]\nonumber\\
&+O\!\left(\frac{r^2\phi^4}{k_{\min}}\right).
\end{align}
At fixed cold-bath temperature, energy gap, and $r\phi^2$,
the leading dependence on the hot-bath temperature vanishes
when $p_{0,*}^{\rm tape}=1-\gamma^\uparrow_{\rm c}
=1/(1+e^{-\beta_c\Delta E})$, or equivalently $\delta_{\rm tape}^*=\beta_c\Delta E$. At this point, 
\begin{equation}
\mathrm{Var}(J_{\rm tape})|_{p^{\rm tape}_0 = p^{\rm tape}_{0,*}
}
 \simeq r\phi^2\gamma^\uparrow_{\rm c}(1-\gamma^\uparrow_{\rm c}).
\end{equation}
Indeed, since
$\partial_{\beta_{\rm h}}\gamma^\uparrow_{\rm h}=-\Delta E\,\gamma^\uparrow_{\rm h}(1-\gamma^\uparrow_{\rm c})$,
the leading derivative changes sign at
$p_0^{\rm tape}=p_{0,*}^{\rm tape}$, explaining the
reversed ordering of the variance curves in Fig.~\ref{fig:current_fluctuations}.

\section{Symmetry of efficiency fluctuations}
\label{app:efficiency_fluctuations}

For stationary, time-symmetric machines, the identification of the reversible efficiency as the least probable one follows from the joint fluctuation symmetry \cite{Verley2014a,Verley2014b}:
\begin{equation}
I(X_{\rm in},X_{\rm out})-I(-X_{\rm in},-X_{\rm out})
=-\eta_{\rm C}X_{\rm in}-X_{\rm out} ,
\label{eq:eraser_required_symmetry}
\end{equation}
where $X_{\rm in}$ represents the current playing the role of the input resource, and $X_{\rm out}$ the useful output. 
In our case, for the eraser regime, the useful output is the accumulated changes in the entropy of the tape $\Delta s_{\rm tape}$ and the input resocurce is the heat current from the hot bath $q_{\rm h}$.

However, the entropy changes $\Delta s_{\rm tape}$ does not generally have the required time-reversal symmetry. Indeed, the increments associated with opposite bit changes are
\begin{equation}
\Delta s_{0\to1}=\ln\frac{p_0^{\rm tape}}{p_1^{\prime\,\rm tape}},
\qquad
\Delta s_{1\to0}=\ln\frac{p_1^{\rm tape}}{p_0^{\prime\,\rm tape}},
\label{eq:eraser_entropy_increments}
\end{equation}
and therefore $\Delta s_{0 \to 1} \neq - \Delta s_{1 \to 0} $. Moreover, an unchanged bit contributes
$\Delta s_{x\to x}=\ln(p_x^{\rm tape}/p_x^{\prime\,\rm tape})$, which
does not need to vanish. As a consequence accumulated stochastic changes in entropy of the tape $\Delta s_{\rm tape}(t)$ is not an antisymmetric current, unlike the accumulated probability current $X_{\rm tape}$ (see Appendix~\ref{app:current_statistics}).
Then, the symmetry in Eq.~\eqref{eq:eraser_required_symmetry} is not guaranteed in this case, and therefore the reversible efficiency need not be the least probable, as illustrated by the efficiency rate functions in the main text.
At the equilibrium point however, since $p^{\prime\,\rm tape}=p^{\rm tape}$, the time-reversal symmetry is recovered, although all average currents vanish.

\section{Quantum repeated-interaction dynamics}
\label{app:coherent_dynamics}
In this appendix, we derive the dynamics of the quantum version of the informational machine using a stochastic repeated-interaction framework~\cite{Manzano2016,Strasberg2017,manzano2018entropy,Manzano2019,Hammam2021}. In this framework, we follow the state of the machine through successive collisions with tape qubits and determine how each interaction modifies both the state of the machine and of the outgoing tape qubit. 

Our starting assumption is that the machine and every incoming qubit are uncorrelated before the collision. Their interaction is then described in the interaction picture as:
\begin{equation}
 \rho_{{\rm mq}}^\prime
 =U(\rho_{\rm m}\otimes\rho_{\rm tape})U^\dagger,
 \qquad U=e^{-iH_{\rm int}\tau},
 \label{eq:app_joint_collision}
\end{equation}
where $\tau$ is the interaction time. Tracing over the tape defines a completely positive and trace preserving (CPTP) map on the machine, $\rho_{\rm m}^\prime = \mathcal{E}(\rho_m)$. Expanding this map through second order in the collision strength $\phi\ll 1$ gives
\begin{align}  \label{eq:app_collision_expansion}
 \mathcal{E}(\rho_{\rm m})={}&\rho_{\rm m}-ic\phi[\sigma_{\rm c}^-\sigma_{\rm h}^++\sigma_{\rm c}^+\sigma_{\rm h}^-,\rho_{\rm m}]+ \\
 &\phi^2 \left( p^{\rm tape}_0\mathcal D[\sigma_{\rm c}^-\sigma_{\rm h}^+]\rho_{\rm m} +
  p^{\rm tape}_1\mathcal D[\sigma_{\rm c}^+\sigma_{\rm h}^-]\rho_{\rm m} \right), \nonumber
\end{align}
where $\mathcal D[L]\rho=L\rho L^\dagger-\{L^\dagger L,\rho\}/2$ and $\sigma_{\rm c}^-\sigma_{\rm h}^+ =\ket{01}\bra{10}$ are dissipators, and we denoted $\sigma_i^- \equiv \ket{0}\!\bra{1}$ and $\sigma_i^+=(\sigma_i^-)^\dagger$, the lowering and raising operators of the qubit, respectively.  As can be observed, in the quantum model, we obtain two different contributions from the interactions with the tape: a first Hamiltonian contribution at first order in $\phi$ due to the coherence $c$ in the tape, and the incoherent transitions proportional to the tape populations at second order in $\phi$, also present in the classical model. Therefore, for $c=0$, the Hamiltonian contribution vanishes and the collision reduces to the classical population dynamics.

We assume that the interaction time of the collisions $\tau$ is fast enough compared with the bath relaxation timescales, so that the action of the baths during the interaction between the tape and the machine is negligible. However, between one collision and the next, the hot and cold baths also act locally on the machine through a thermal generator in Lindblad form:
\begin{equation}
 \mathcal L_0\rho_{\rm m}=
 \sum_{i=c,h}\left(
 \Gamma_i^\uparrow\mathcal D[\sigma_i^+]\rho_{\rm m}
 +\Gamma_i^\downarrow\mathcal D[\sigma_i^-]\rho_{\rm m}
 \right),
 \label{eq:app_thermal_generator}
\end{equation}
where $\sigma_i^\pm$ are the raising and lowering operators of subsystem $i$. To formally combine the continuous thermal evolution with the collisions, let us denote $\rho_{\rm m}^{(n)}$ the unnormalized machine state conditioned on exactly $n$ collisions having occurred. Its trace is the probability of that collision count. Assuming random and uncorrelated arrival times at rate $r$ (Poisson statistics) and fast interaction times $\tau$, these conditional states obey
\begin{equation}
 \dot\rho_{\rm m}^{(n)}
 =(\mathcal L_0-r\mathbb I)\rho_{\rm m}^{(n)}
  +r\mathcal{E}(\rho_{\rm m}^{(n-1)}),
 \label{eq:app_poisson_recursion}
\end{equation}
with $\rho_{\rm m}^{(-1)}=0$.  Summing over $n$ removes the conditioning in Eq.~\eqref{eq:app_poisson_recursion} and yields $\dot\rho_{\rm m}=\mathcal L_0\rho_{\rm m}+r[ \mathcal{E}(\rho_{\rm m})-\rho_{\rm m}]$. Finally, using Eq.~\eqref{eq:app_collision_expansion}, the resulting master equation is
\begin{align}
 \dot\rho_{\rm m}={}&-i[V,\rho_{\rm m}]+\mathcal L_0\rho_{\rm m}
 \nonumber\\
 &+\Gamma_{\rm tape}^\uparrow\mathcal D[\sigma_{\rm c}^-\sigma_{\rm h}^+]\rho_{\rm m}
  +\Gamma_{\rm tape}^\downarrow\mathcal D[\sigma_{\rm c}^+\sigma_{\rm h}^-]\rho_{\rm m}.
 \label{eq:app_coherent_master}
\end{align}
We notice that the Hamiltonian contribution $V$ only couples the degenerate states of the machine $\ket{01}_{\rm m}$ and $\ket{10}_{\rm m}$. Accordingly, the only coherence sustained in the stationary state is in this subspace, while the remaining coherences decay in the long time run due to the thermal dissipation. In the ordered basis $\{\ket{00}_{\rm m},\ket{01}_{\rm m},\ket{10}_{\rm m},\ket{11}_{\rm m}\}$, the stationary state reads
takes the form
\begin{equation}
 \rho_{\rm m}^{\rm ss}=
 \begin{pmatrix}
 p_{00}^{\rm ss}&0&0&0\\
 0&p_{01}^{\rm ss}&z^{\rm ss}&0\\
 0&z^{{\rm ss}*}&p_{10}^{\rm ss}&0\\
 0&0&0&p_{11}^{\rm ss}
 \end{pmatrix}.
 \label{eq:app_stationary_density}
\end{equation}
where the specific values of $p_{i j}^{\rm ss}$ and $z_{\rm ss}$ are given in Ref.[].
The transformation of the tape can be obtained by tracing the machine in Eq.~\eqref{eq:app_joint_collision}. Taking the stationary regime in the machine this defines a CPTP map on the tape qubits $\rho_{\rm tape}' = \mathcal A(\rho_{\rm tape})$, whose expansion to second order in $\phi$ reads
\begin{align}
 \mathcal A(\rho_{\rm tape}) =& \rho_{\rm tape}-i\phi[z^{\rm ss}\sigma_{\rm tape}^+ + z^{{\rm ss}*}\sigma_{\rm tape}^ -,\rho_{\rm tape}]
 \nonumber\\
 &+\phi^2\bigl(
 p_{01}^{\rm ss}\mathcal D[\sigma_{\rm tape}^+]
 +p_{10}^{\rm ss}\mathcal D[\sigma_{\rm tape}^-]
 \bigr)\rho_{\rm tape}.
 \label{eq:app_outgoing_tape_coherent}
 \end{align}
 The stationary machine coherence induces a Hamiltonian term, while the machine populations lead to incoherent jumps between the states $\ket{0}$ and $\ket{1}$ of the tape qubit.
Using the spectral decomposition of the tape qubits, $\rho_{\rm tape}=\sum_{k=\pm}\lambda_k \Pi_k$ with 
$\Pi_k \equiv \ket{\lambda_k}\!\bra{\lambda_k}$, we expand the eigenvalues of $\rho_{\rm tape}^\prime$ as
$\lambda_k'=\lambda_k+\phi\lambda_k^{(1)} +\phi^2\lambda_k^{(2)}+O(\phi^3)$.
Since the first-order commutator in Eq.~\eqref{eq:app_outgoing_tape_coherent} describes an infinitesimal
unitary rotation, it has no contribution to the diagonal elements of $\rho_{\rm tape}^\prime$. Consequently, $\lambda_k^{(1)}=0$, and the leading contribution appears at second order:
\begin{align}
& \lambda_k^{(2)}=
 p_{01}^{\rm ss}
 \langle\lambda_k|\mathcal D[\sigma_q^+]
 \rho_{\rm tape}|\lambda_k\rangle
 +p_{10}^{\rm ss}
 \langle\lambda_k|\mathcal D[\sigma_q^-]
 \rho_{\rm tape}|\lambda_k\rangle
 \nonumber\\
 &+\sum_{l\ne k}(\lambda_k-\lambda_l)
 |\langle\lambda_l|z^{\rm ss}\sigma_{\rm tape}^+ + z^{{\rm ss}*}\sigma_{\rm tape}^ -|\lambda_k\rangle|^2,
 \label{eq:app_tape_eigenvalue_correction}
\end{align}
where the first two terms come from the incoherent part of the collision map, and the last term is induced by the first-order correction to the eigenvectors. Normalization implies $\sum_k\lambda_k^{(2)}=0$. 

The von Neumann entropy changes and tape-induced probability fluxes are calculated from Eq.~\eqref{eq:app_tape_eigenvalue_correction} as
\begin{equation}
 \begin{split}
 \dot S_{\rm tape}
 &=-r\phi^2\sum_{k=\pm}\lambda_k^{(2)}\ln\lambda_k, \\
 J_{\rm tape}^{\rm q} &= \frac r2\operatorname{Tr}
 [(\rho_{\rm tape}'-\rho_{\rm tape})(\Pi_+-\Pi_-)].
 \label{eq:app_quantum_entropy_perturbation}
\end{split}
\end{equation}
Since $\ln\rho_{\rm tape} =\tfrac12\ln(\lambda_+\lambda_-)\mathbb 1
+\tfrac12 \delta_{\rm tape}^q(\Pi_+-\Pi_-)$, trace preservation and the definition of quantum relative entropy gives
$-\delta_{\rm tape}^q J_{\rm tape}^q =\dot S_{\rm tape} +rD(\rho_{\rm tape}'\Vert\rho_{\rm tape})$, as in the classical case.

For computing the FCS statistics we introduce the tilted dissipator associated with bath $i$ is
\begin{align} \label{eq:quantum_tilted_dissipator}
\mathcal D_{i,\lambda_i}(\rho_{\rm m})
=~ &\Gamma_i^\uparrow\left[
e^{\lambda_i\Delta E}\sigma_i^+\rho_{\rm m}\sigma_i^-
-\frac12\{\sigma_i^-\sigma_i^+,\rho_{\rm m}\}
\right] \nonumber \\
+&\Gamma_i^\downarrow\left[
e^{-\lambda_i\Delta E}\sigma_i^-\rho_{\rm m}\sigma_i^+
-\frac12\{\sigma_i^+\sigma_i^-,\rho_{\rm m}\}
\right]. \nonumber
\end{align}
where $\lambda_i$ for $i={\rm h, c}$ are counting fields associated to the thermal baths. The full tilted Liouvillian in this case reads
\begin{align} 
\mathcal L_{\boldsymbol\lambda}\rho_{\rm m}
={}&-i[V,\rho_{\rm m}]
+\sum_{i=h,c}\mathcal D_{i,\lambda_i}(\rho_{\rm m}) 
\\
&+\Gamma_{\rm tape}^\uparrow\mathcal D[\sigma_{\rm c}^-\sigma_{\rm h}^+]\rho_{\rm m}
+\Gamma_{\rm tape}^\downarrow\mathcal D[\sigma_{\rm c}^+\sigma_{\rm h}^-]\rho_{\rm m}, \nonumber
\end{align}
where we did not included counting in the tape qubits. The auxiliary operator $\rho_{\rm m, \boldsymbol\lambda}(t)$ obeys
$\dot\rho_{\rm m,\boldsymbol\lambda}
=\mathcal L_{\boldsymbol\lambda}\rho_{\rm m,\boldsymbol\lambda}$,
with $\rho_{\boldsymbol\lambda}(0)=\rho_{\rm m}(0)$.
Its trace generates the joint heat statistics $
\operatorname{Tr}[\rho_{\rm m,\boldsymbol\lambda}(t)]
=
\left\langle
e^{\lambda_h q_h(t)+\lambda_c q_c(t)},
\right\rangle,$

and the corresponding scaled cumulant generating function is $\Psi(\boldsymbol\lambda) = \lim_{t\to\infty}\frac1t \ln  \operatorname{Tr}\rho_{\rm m,\boldsymbol\lambda}(t)$.

After vectorizing the density operator, we denote the matrix
representation of $\mathcal L_{\boldsymbol\lambda}$ by
$\mathbb W_{\boldsymbol\lambda}$. The SCGF is its dominant
eigenvalue, continuously connected to the stationary
eigenvalue $\Psi(\boldsymbol0)=0$. Here
$\mathbb W_{\boldsymbol\lambda}$ acts in Liouville space
and retains the populations and the coherences coupled
to them. To evaluate the stationary heat currents and their asymptotic variance rate without explicitly solving the eigenvalue problem, we use inverse counting statistics~\cite{Bruderer2014}. The characteristic polynomial of $\mathbb{W}_{\mathbb \lambda}$ reads
\begin{equation}
P(z,\boldsymbol\lambda)
=
\det\!\left[z\mathbb I-\mathbb W_{\boldsymbol\lambda}\right]
=
\sum_{\mu=0}^{6}a_\mu(\boldsymbol\lambda)z^\mu,
\label{eq:inverse_counting_polynomial}
\end{equation}
and the eigenvalue equation implies $P\bigl(\Psi(\boldsymbol\lambda),\boldsymbol\lambda\bigr)=0$. Taking the first two derivatives with respect to $\lambda_i$ and evaluating them at zero fields yields
\begin{align}
\dot Q_i
&=-\frac{a_0'}{a_1},
\nonumber\\
\operatorname{Var}(\dot Q_i)
&=-\frac{
a_0''+2a_1'\dot Q_i+2a_2(\dot Q_i)^2
}{a_1},
\label{eq:inverse_counting_cumulants}
\end{align}
where we denoted $a_\mu=a_\mu(\boldsymbol0)$, $a_\mu'=\left.\partial_{\lambda_i}a_\mu\right|_{\boldsymbol0}$, and 
$a_\mu''=\left.\partial_{\lambda_i}^2a_\mu\right|_{\boldsymbol0}$ for $i= {\rm h,c}$.

\bibliography{biblio}
\end{document}